\documentclass[lettersize,journal]{IEEEtran}
\usepackage{amsmath,amsfonts,amssymb}
\usepackage{bm}
\usepackage{algorithm}
\usepackage{algorithmic}
\usepackage{array}
\usepackage{textcomp}
\usepackage{stfloats}
\usepackage{url}
\usepackage{verbatim}
\usepackage{graphicx}
\usepackage{caption}
\usepackage{cite}
\ifCLASSOPTIONcompsoc
  \usepackage[caption=false,font=normalsize,labelfont=sf,textfont=sf]{subfig}
\else
  \usepackage[caption=false,font=footnotesize]{subfig}
\fi

\begin{document}

\title{Opportunistic Routing in Wireless Communications via Learnable State-Augmented Policies}

\author{Sourajit Das, Kirtan Gopal Panda, Navid NaderiAlizadeh

\thanks{This work will be presented in part at the 2025 IEEE International Conference on Acoustics, Speech and Signal Processing (ICASSP) \cite{das2025icassp}.}
}



\maketitle

\begin{abstract}
This paper addresses the challenge of packet-based information routing in large-scale wireless communication networks. The problem is framed as a constrained statistical learning task, where each network node operates using only local information. Opportunistic routing exploits the broadcast nature of wireless communication to dynamically select optimal forwarding nodes, enabling the information to reach the destination through multiple relay nodes simultaneously. To solve this, we propose a State-Augmentation (SA) based distributed optimization approach aimed at maximizing the total information handled by the source nodes in the network. The problem formulation leverages Graph Neural Networks (GNNs), which perform graph convolutions based on the topological connections between network nodes. Using an unsupervised learning paradigm, we extract routing policies from the GNN architecture, enabling optimal decisions for source nodes across various flows. Numerical experiments demonstrate that the proposed method achieves superior performance when training a GNN-parameterized model, particularly when compared to baseline algorithms. Additionally, applying the method to real-world network topologies and wireless ad-hoc network test beds validates its effectiveness, highlighting the robustness and transferability of GNNs.
\end{abstract}


\section{Introduction}

\IEEEPARstart{T}{he} rapid growth of digital communications and high-speed Internet has resulted in the prolific use of wireless devices and smart systems, triggering both advancements and challenges in fields like IoT, autonomous transportation, and massive MIMO systems that require more bandwidth, greater coverage, and improved reliability. Traditional networks persistently face scalability, robustness, and quality of service (QoS) issues which can now be addressed through Artificial Intelligence (AI). More specifically machine learning (ML), has been instrumental in solving these modern day communication problems by providing solutions that were previously inaccessible ~\cite{mao2018deep}. As these wireless systems are incorporated into smart infrastructures, they must cope with increased traffic, unpredictable conditions, and risks of transmission delays and packet loss ~\cite{rozner2009soar}. Routing is a key solution to improve the quality of service in which nodes in the network exchange information packets using communication protocols based on certain standardized protocols ~\cite{MEDHI201864}. 

In recent times, opportunistic routing (OR) has become prevalent in wireless networks by leveraging the broadcast nature of the medium. In contrast to conventional methods that depend on a fixed, predefined path, OR facilitates
multiple routes, thereby ensuring both efficiency and reliability of operation \cite{biswas2005ExOR}. This advantage becomes particularly evident in scenarios with highly mobile infrastructure or frequent link disruptions, just like in vehicular ad-hoc networks (VANETs) and mobile ad-hoc networks (MANETs). Most OR protocols rely on a flooding scheme, necessitating intelligent path selection to minimize overhead expenses. Additionally, frequent changes in network topologies due to motion and uncertainties in parameters complicate the process, often increasing the processing overhead due to control messages~\cite{kaviani2023deepmpr}. Deep learning techniques help address these complexities by predicting network dynamics amidst the uncertainties. In this study, we solve the challenge of joint routing along with scheduling in packet based opportunistic networks by leveraging the power of Graph Neural Networks (GNN).

Communication networks incorporate advanced mechanisms such as radio resource allocation, congestion management, and queue management, which significantly affect system performance. Numerous studies explored these challenges through stochastic network utility maximization (NUM) \cite{liu2015joint}, traffic engineering and routing \cite{xia2018utility, zargham2013accelerated, das2023learning}, radio resource management (RRM) \cite{eisen2020optimal, naderializadeh2023learning, naderializadeh2022state}, and link scheduling \cite{xia2017stochastic, xia2014distributed} 
These works typically model the problems as constrained optimization tasks involving a utility function while considering the stochastic nature of user traffic and variations in wireless channel conditions. Mao et al. provide an in-depth analysis of how machine learning (ML) techniques can enhance network operations, including resource allocation, path optimization, traffic management, and data compression \cite{mao2018deep}. Of late supervised learning has proven effective at mimicking system heuristics using training datasets ~\cite{sun2017learning, xu2019energy, van2019sum}, unfortunately it requires constant data collection and may not always surpass heuristic approaches. 
In contrast, this paper employs unsupervised learning, treating the network optimization as a statistical regression problem to solve the problem directly without depending on training sets, which can outperform heuristic solutions \cite{eisen2020optimal,de2018team,meng2020power,  cui2019spatial}. 
Earlier, fully connected neural networks (FCNN) were widely adopted due to their universal approximation ability \cite{sun2017learning, eisen2020optimal},
followed by convolutional neural networks (CNN) for their scalability in time and space. However, CNNs struggle on large-scale networks and lack generalization due to permutation invariance \cite{zhang2023admire, he2020machine, shen2022graph}. This motivates us to use graph neural networks (GNN) for communication scenarios, which offer scalability, portability and improved performance \cite{bernardez2023magnneto, henaff2015deep, gama2018convolutional}. 

This paper addresses the challenge of network utility maximization under multiple constraints, drawing inspiration from methods presented in prior studies ~\cite{eisen2020optimal, naderializadeh2023learning, naderializadeh2022state, xia2018utility, zargham2013accelerated, uslu2024learning, xia2014distributed}. Our focus is on improving packet-based routing and scheduling to enhance the average performance of network nodes, ensuring stable queue lengths over time and maintaining node stability. The optimization framework requires strict adherence to constraints to preserve system feasibility. These types of problems are often addressed by transitioning to the Lagrangian domain, in which one single objective function is optimized with respect to primal variables while being minimized over dual variables. The primal variables are associated with the main objective, whereas the dual variables correspond to the constraints of the network optimization problem. Although primal-dual methods can theoretically provide optimal solutions, challenges such as residual duality gaps remain significant obstacles.

To address these challenges, we propose a constrained learning approach based on state augmentation, which incorporates dual variables to capture the extent of constraint satisfaction or violation over time~\cite{calvo2021state}. In this framework, we enhance the standard network state by appending dual multipliers at every time step, using them as continuous inputs for the network routing policy. This integration of dual multipliers into the network routing policy facilitates the model to adjust its decisions in response to real-time channel conditions while ensuring that the system adheres to the imposed constraints. Unlike our previous work in ~\cite{das2023learning} which was limited to wired networks, we focus our attention on wireless communication while realizing them on actual networks. We can summarize our contributions to the paper as below:
\begin{itemize}
    \item We perform opportunistic routing using Graph Neural Networks to learn efficient routing strategies for wireless communication networks of varying sizes and validated them on real time wireless ad-hoc networks.
    \item The Graph Neural Networks (GNNs) enables us to implement decentralized communication networks of varying sizes by mimicking the Method of Multipliers (MoM) which is faster than standard dual descent methods.
    \item We utilize State Augmentation methods to learn near-optimal solutions with a finite number of iterations unlike other iterative algorithms which take a longer time to converge.
\end{itemize}


The structure of this paper is designed as follows: Section \ref{sec:problem} introduces the problem, where we frame it as a network utility optimization problem subject to multiple constraints. Sections \ref{sec:dual_desc} and \ref{sec:mom} provide an overview of standard optimization techniques. Section \ref{sec:gnn_par} details the process of parameterizing the learning algorithm using Graph Neural Networks (GNNs). Section \ref{sec:gnn_mom} explores the key concepts of the augmented Lagrangian method and GNN learning, highlighting their relevance to the problem. Section \ref{sec:state_aug} outlines the proposed state-augmented algorithm for optimizing routing decisions in the network. Section \ref{sec:results} evaluates and compares the performance of various optimization methods. Finally, the paper concludes with key insights in Section \ref{sec:conclusion}.

\section{Problem Statement} \label{sec:problem}


Let's represent the communication network state as a graph $\mathcal{G} = (\mathcal{V, E})$. Here $\mathcal{V}$ denotes the set of communicating nodes, while $\mathcal{E} \subseteq \mathcal{V} \times \mathcal{V}$ represents the edges connecting these nodes. The neighborhood of a node $i$ is defined as $\mathcal{N}_i = \{j \in \mathcal{V} | (i,j) \in \mathcal{E} \}$, which includes all nodes $j$ that have direct communication links with $i$. Information packets are exchanged between nodes across various flows, with $\mathcal{K}$ denoting the set of flows. Each flow $k \in \mathcal{K}$ has a designated destination node $o_k \in \mathcal{V}$. For a particular time instant $t$, a node $i\neq o_k $ produces a certain number of random packets, $a_{0i}^k(t)$, intended for the destination $o_k$. These random variables $a_{0i}^k(t)$ are assumed to be independent and identically distributed (\textit{iid}) over time, with their mean value given by $\mathbb{E}[a_{0i}^k(t)]=a_{0i}^k$. Considering the link state conditions, the probability that node $i$ successfully decodes a packet sent by node $j$ is represented by $\mathrm{R}_{ij}$, an element of the channel state probability matrix $\mathbf{R} \in \mathbb{R}^{|\mathcal{V}| \times |\mathcal{V}|}$, which captures the channel state probability between pairs of communicating nodes.

Since we perform OR by taking advantage of the broadcast nature of the wireless medium, we consider two different routing variables. We consider the first routing variable $\mathrm{T}_i^k$, to denote the probability that node $i$ in the network decides to transmit a packet to its neighbors over flow $k$. Once the broadcasted packets arrive at node $i$ for flow $k$, the probability that node $i$ decides to keep the packet from node $j$ be denoted using the second routing variable as $\mathrm{K}_{ij}^k$. Thus at a given time instant and flow $k$, the number of packets received from node $j$ to node $i$ can be given as the product of the above probabilities times the number of packets generated at node $j$, which is $\mathrm{T}_j^k(t) ~ \mathrm{R}_{ij}(t) ~ \mathrm{K}_{ij}^k(t) ~ a_{0j}^k(t)$.
Each node $i$ can transmit a fixed number of packets based on its capacity, denoted as $\mathrm{C}_i$. The balance between the total packets received, which include $a_{0i}^k(t)$ and  $\sum_{j \in n_i} \mathrm{T}_j^k(t) ~ \mathrm{R}_{ij}(t) ~ \mathrm{K}_{ij}^k(t) ~ a_{0j}^k(t)$, and the total packets transmitted, $\mathrm{T}_i^k(t) ~ \mathrm{C}_i$, determines the change in the local queue. If the balance is positive, it is added to the queue; if negative, packets are subtracted. Consequently, the queue length for flow $k$ at node $i$, denoted by $q_i^k(t)$, evolves iteratively based on the equation below:
\begin{multline} \label{queueupdate}
    q_i^k(t+1) = \Bigr[q_i^k(t) + a_{0i}^k(t) + \\
    \sum_{j \in n_i} \mathrm{T}_j^k(t) ~ \mathrm{R}_{ij}(t) ~ \mathrm{K}_{ij}^k(t) ~ a_{0j}^k(t) - \mathrm{T}_i^k(t) ~ \mathrm{C}_i \Bigr]^+.  
\end{multline}
Here, the projection onto the non-negative orthant ensures that the queue length remains non-negative. It is important to note that this eq. \eqref{queueupdate} applies to all nodes $i\neq o_k $, because the packets that reach their destination $o_k$ are removed from the system.

\subsection{Problem Design}
\label{ssec:problem_subhead}
This study examines a communication network over a sequence of time instants $t \in \{0, 1,..,T-1 \}$. At each time instant $t$, let's denote the network state, specifically the set of channel probabilities, as $\mathbf{R}_t \in \mathcal{R}$. For any particular network state, let us denote the routing decision vectors as $\mathbf{p}(\mathbf{R}_t)$ and $\mathbf{b}(\mathbf{R}_t)$. Here $\mathbf{p}:\mathcal{R} \rightarrow \mathbb{R}^{n \times n \times F}$ maps the network state to routing decisions regarding acceptance of the packets at a given node. On the other hand, $\mathbf{b}:\mathcal{R} \rightarrow \mathbb{R}^{n \times F}$ maps the network state to routing decisions regarding the transmission of packets from a particular node. These routing decisions influence the network's performance which can be expressed as a vector $\mathbf{f}(\mathbf{R}_t, \mathbf{p}(\mathbf{R}_t), \mathbf{b}(\mathbf{R}_t)) \in \mathbb{R}^b$, with $\mathbf{f}:\mathcal{R} \times \mathbb{R}^{n \times n \times F} \times \mathbb{R}^{n \times F} \rightarrow \mathbb{R}^b$ being the network's performance function.


An auxiliary optimization variable, let's say $a_i^k(t) \geq a_{0i}^k(t)$ is introduced to maximize packet generation for flow $k$ at node $i$. Here, $a_{0i}^k(t)$ represents the true number of packets in the network at time $t$, which is essential for updating the length of the queues in the network, as described in eq. \eqref{queueupdate}. Following the general framework in \cite{naderializadeh2022state, das2023learning}, ``a concave utility function'', $\mathcal{U}: \mathbb{R}^x \rightarrow \mathbb{R}$ along with a given number of constraints $\mathbf{g}: \mathbb{R}^x \rightarrow \mathbb{R}^y$ are considered. The network routing formulation is now formulated as:
\begin{subequations} \label{eq:concave_util}
\begin{align}
    \max_{[\mathbf{p}(\mathbf{R}_t), \mathbf{b}(\mathbf{R}_t)]_{t=0}^{T-1}} \hspace{0.5cm} \mathcal{U} \Biggl(\frac{1}{T} \sum_{t=0}^{T-1} \mathbf{f} \Bigl(\mathbf{R}_t, \mathbf{p}(\mathbf{R}_t), \mathbf{b}(\mathbf{R}_t) \Bigr)\Biggr) \\
    s.t. \hspace{1cm} \mathbf{g}\Biggl(\frac{1}{T} \sum_{t=0}^{T-1} \mathbf{f}\Bigl(\mathbf{R}_t, \mathbf{p}(\mathbf{R}_t), \mathbf{b}(\mathbf{R}_t)\Bigr) \Biggr) & \geq 0.
\end{align}
\end{subequations}
Here both the objective function as well as the set of constraints are established on the \textit{ergodic average} of the system performance; $\frac{1}{T} \sum_{t=0}^{T-1} \mathbf{f}(\mathbf{R}_t, \mathbf{p}(\mathbf{R}_t), \mathbf{b}(\mathbf{R}_t))$. Thus the goal of the network routing algorithm is to find out the optimal routing decision vectors $\mathbf{p}(\mathbf{R}_t)$ and $\mathbf{b}(\mathbf{R}_t)$ for any given network state $\mathbf{R}_t \in \mathcal{R}$. 

In order to address the network optimization problem at hand, we begin from the network utility function which aims to maximize the information packets at every node $i$ and across all flows $k$. Now we can formulate the concave objective function as: 
\begin{multline} \label{eq:util_function}
    \mathcal{U} \Biggl(\frac{1}{T} \sum_{t=0}^{T-1} \mathbf{f}(\mathbf{R}_t, \mathbf{p}(\mathbf{R}_t), \mathbf{b}(\mathbf{R}_t))\Biggl) = \\
    \sum_{k \in \mathcal{K}} \sum_{i \in \mathcal{V}} \log \Biggl(\frac{1}{T} \sum_{t=0}^{T-1} a_i^k(t) \Biggl).
\end{multline}

The optimization problem includes three types of constraints based on the current network state:

\begin{enumerate}
    \item \textbf{Routing Constraints:} The total packets received and local packets $a_i^k(t)$ at node $i$ must not exceed the total transmitted packets from node $i$:
    \begin{equation} \label{eq:flow_constr}
    a_i^k(t) + \sum_{j \in n_i} \mathrm{T}_j^k(t) ~ \mathrm{R}_{ij}(t) ~ \mathrm{K}_{ij}^k(t) ~ a_j^k(t) \leq \mathrm{T}_i^k(t) ~ \mathrm{C}_i .
\end{equation}

    \item \textbf{Minimum Constraints:} Each node $i$ must maintain a minimum number of local packets for faithful transmission to other communicating nodes in the network:
    \begin{equation} \label{eq:min_constr}
    a_i^k(t) \geq a_{0i}^k(t).
\end{equation}

    \item \textbf{Capacity Constraints:} This defines the maximum probability of transmitting packets from a node, \textit{i.e.} the total sum of probabilities of transmitted packets from node $i$ can \emph{not} exceed the maximum probability which is 1:
    \begin{equation} \label{eq:cap_constr}
    \sum_{k \in K} T_i^k(t) \leq 1.
\end{equation}
\end{enumerate}


By incorporating these constraints \eqref{eq:util_function}-\eqref{eq:cap_constr} into the defined formula in~\eqref{eq:concave_util}, the network utility maximization problem can be expressed below:
\begin{subequations} \label{eq:prob_state}
\begin{align}
    \max_{[a_i^k(t), \mathrm{T}_i^k(t), K_{ij}^k(t)]_{t=0}^{T-1}} \hspace{0.25cm}  \sum_{k \in \mathcal{K}} \sum_{i \in \mathcal{V}} \log \Biggl(\frac{1}{T} \sum_{t=0}^{T-1} a_i^k(t)\Biggl) \label{eq:7a} \\
    \mathrm{T}_i^k(t) ~ \mathrm{C}_i - a_i^k(t) - \sum_{j \in n_i} \mathrm{T}_j^k(t) ~ \mathrm{R}_{ij}(t) ~ \mathrm{K}_{ij}^k(t) ~ a_j^k(t) \geq 0 \label{eq:7b} \\
    a_i^k(t) - a_{0i}^k(t) \geq 0 \label{eq:7c} \\
    \sum_{k \in K} \mathrm{T}_i^k(t) \leq 1 \label{eq:7d} 
\end{align}
\end{subequations}

\section{Gradient Based Solutions for Network Routing: Dual Domain} \label{sec:dual_desc}

Given that the network maximization challenge in ~\eqref{eq:prob_state} consists of a concave utility function, the standardized gradient-based dual descent algorithm is a desirable approach. We consider a Lagraingian dual multiplier, $\bm{\mu} \in \mathbb{R}_+^c$, associated with the constraint \eqref{eq:7b}. Furthermore, the constraints \eqref{eq:7c}, \eqref{eq:7d} are kept implicit, and for simplicity in our analysis, we set the total time, $T = 1$. The Lagrangian can now be formulated as:
\begin{multline} \label{eq:dd_lagangian}
    \mathcal{L} (\bm{a}, \mathbf{T, K}, \bm{\mu}) = \sum_i \sum_k \log ( a_i^k ) \\
    + \sum_i \sum_k \mu_i^k \Bigg( \mathrm{T}_i^k ~ \mathrm{C}_i - a_i^k - \sum_{j \in n_i} \mathrm{T}_j^k ~ \mathrm{R}_{ij} ~ \mathrm{K}_{ij}^k ~ a_j^k \Bigg).
\end{multline}
We can maximize the above Lagrangian in \eqref{eq:dd_lagangian} using a standardized gradient-based algorithm. Specifically, a 
primal-dual learning algorithm, a variant of dual-descent algorithm, aims at maximizing $\mathcal{L}$ over the primal variables $\bm{a}$, $\mathbf{T}$ and $\mathbf{K}$, while minimizing over the dual multipliers $\bm{\mu}$ concurrently.
\begin{equation} \label{eq:dd_lag_update}
    \mathcal{L}^* = \min_{\bm{\mu}} \max_{\bm{a},\mathbf{T, K}} \hspace{0.25cm} \mathcal{L}(\bm{a}, \mathbf{T, K}, \bm{\mu})
\end{equation}
For each iteration $m \in \{1,2,..,M\}$, the primary variables are updated as per the following equations in \eqref{eq:dd_update},
\begin{subequations} \label{eq:dd_update}
\begin{align} 
    [{a_i^k}]_{m+1} = [{a_i^k}]_{m} + \gamma_{\bm{\phi}} \bm{\nabla}_{\phi} \mathcal{L} (\bm{a}, \mathbf{T, K}, \bm{\mu}) \label{eq:dd_update_a} \\
    [{\mathrm{T}_i^k}]_{m+1} = [{\mathrm{T}_i^k}]_{m} + \gamma_{\bm{\phi}} \bm{\nabla}_{\phi} \mathcal{L} (\bm{a}, \mathbf{T, K}, \bm{\mu}) \label{eq:dd_update_T} \\
    [{K_{ij}^k}]_{m+1} = [{K_{ij}^k}]_m + \gamma_{\bm{\phi}} \bm{\nabla}_{\phi} \mathcal{L} (\bm{a}, \mathbf{T, K}, \bm{\mu}) \label{eq:dd_update_K}
\end{align}
\end{subequations}
The gradient descent update of the dual variable is given by $\bm{\mu}_{m+1} =  \bm{\mu}_m - \gamma_{\bm{\mu}} \bm{\nabla}_{\phi} \mathcal{L} (\bm{a}, \mathbf{T, K}, \bm{\mu})$. Due to the linearity of the Lagrangian with respect tp $\bm{\mu}$, its gradient is much easier to calculate, allowing for recursive updates of $\bm{\mu}$ as,
\begin{multline} \label{dd:dual_update}
    \bm{\mu}_{m+1} = \Bigg[ \bm{\mu}_m - \gamma_{\bm{\mu}} \Big( \mathrm{T}_i^k ~ \mathrm{C}_i - a_i^k - \sum_{j \in n_i} \mathrm{T}_j^k ~ \mathrm{R}_{ij} ~ \mathrm{K}_{ij}^k ~ a_j^k\Big) \Bigg]^+
\end{multline}
Here, $[.]^+$ denotes the non-negative orthant; it is expressed as $[d]^+ = max(d,0)$. The constants $\gamma_{\bm{\mu}}$ and $\gamma_{\bm{\phi}}$ represent the learning rates or step size of the dual variable ($\bm{\mu}$) and the primal variables ($\bm{a}$, $\mathbf{T}$, $\mathbf{K}$), respectively. With the introduction of dual variables in the Lagrangian, this routing algorithm can yield nearly optimal and feasible results for this dual relaxation over an adequate number of iterations. Unlike standard gradient descent methods like primal-dual learning, which may not always produce a feasible set of routing decisions, this method can guarantee feasibility for the routing algorithms in \eqref{eq:dd_update} and \eqref{dd:dual_update}. However, a significant drawback of this dual descent algorithm is its slow convergence rate. To address this issue, we explore the Method of Multipliers in the following section as an alternative approach.

\subsection{Solution via Method of Multipliers (MoM)} \label{sec:mom}
While addressing any network utility optimization within the framework of Lagrangian dual, the dual descent method is often a sensible starting point for finding an optimal solution. However, this method has well documented disadvantages, which includes its very slow rate of convergence and the requirement for a strictly convex objective function \cite{bertsekas2015parallel}. Thus, we resort to the Augmented Lagrangian, also known as Method of Multipliers (MoM), which is a superior alternative for mitigating the shortcomings of the dual descent method similar to our previous work in \cite{das2023learning}. 
The algorithm begins by converting the inequality constraint in \eqref{eq:7b} into an equality constraint, as follows: 
\begin{equation} \label{eq:ineq_to_equal}
    \mathrm{T}_i^k(t) ~ \mathrm{C}_i - a_i^k(t) - \sum_{j \in n_i} \mathrm{T}_j^k(t) ~ \mathrm{R}_{ij}(t) ~ \mathrm{K}_{ij}^k(t) ~ a_j^k(t) - z_i^k(t) = 0,
\end{equation}
where $z_i^k \geq 0$ is an auxiliary variable introduced to help with the inequality. The Augmented Lagrangian of the above network utility optimization in \eqref{eq:prob_state} is hence formulated as:
\begin{multline} \label{aug_lag_mom}
     \mathcal{L}_{\rho} (\bm{a}, \mathbf{T, K},  \bm{z}, \bm{\mu}) = \sum_{k\in\mathcal{K}} \sum_{i\in\mathcal{V}} \frac{1}{T} \sum_{t=0}^{T-1} \log ( a_i^k ) \\
    + \sum_{k\in\mathcal{K}}\sum_{i\in\mathcal{V}} \frac{1}{T} \sum_{t=0}^{T-1} \Bigg\{ \mu_i^k \Bigg( \mathrm{T}_i^k ~ \mathrm{C}_i - a_i^k - \sum_{j \in n_i} \mathrm{T}_j^k ~ \mathrm{R}_{ij} ~ \mathrm{K}_{ij}^k ~ a_j^k - z_i^k \Bigg) \\
    + \frac{\rho}{2} \Biggl|\!\Biggl| \mathrm{T}_i^k ~ \mathrm{C}_i - a_i^k - \sum_{j \in n_i} \mathrm{T}_j^k ~ \mathrm{R}_{ij} ~ \mathrm{K}_{ij}^k ~ a_j^k - z_i^k \Biggr|\!\Biggr|^2 \Bigg\},
\end{multline}
where $\rho > 0$ is the penalty factor while $\bm{\mu} \in \mathbb{R}_+^{n \times F}$ is the dual variable corresponding to the routing constraint in \eqref{eq:7b}. Note that constraints \eqref{eq:7c}, \eqref{eq:7d} are implicitly enforced by ensuring that the solutions of the given optimization problem satisfy the constraints inherently.
This penalty method tackles the above network optimization by iteratively varying the $\bm{\mu}$ and $\rho$.
\begin{subequations} \label{eq:MoM_algo}
\begin{align}
    \bm{a}^{m+1}, \mathbf{T}^{m+1}, \mathbf{K}^{m+1}, \bm{z}^{m+1} = \arg \max_{\bm{a}, \mathbf{T, K}, \bm{z}} \mathcal{L}_{\rho} (\bm{a}, \mathbf{T, K}, \bm{z}, \bm{\mu}^m) \\
    \bm{\mu}^{m+1} = \arg \min_{\bm{\nu}} \mathcal{L}_{\rho} (\bm{a}^{m+1}, \mathbf{T}^{m+1}, \mathbf{K}^{m+1}, \bm{z}^{m+1}, \bm{\mu}^m).\label{eq:mom_min_dual}
    \end{align}
\end{subequations}
As we observe that $\mathcal{L}_{\rho} (\bm{a}, \mathbf{T, K}, \bm{\mu})$ is linear in $\bm{\mu}$, we can perform the minimization step for $\bm{\mu}$ from ~\eqref{eq:mom_min_dual} via the gradient descent:
\begin{multline} \label{eq:mom_dual}
    (\mu_i^k)^{m+1} = \Bigg[(\mu_i^k)^{m} - \rho^m \bigg(\mathrm{T}_i^k ~ \mathrm{C}_i \\ 
    - \sum_{j \in n_i} \mathrm{T}_j^k ~ \mathrm{R}_{ij} ~ \mathrm{K}_{ij}^k ~ a_j^k - (a_i^k)^m \bigg) \Bigg]^+. 
\end{multline}

The MoM algorithm is guaranteed to converge under specific conditions, particularly once the maximization operation across $\bm{z}$ outputs a optimal solution irrespective of its initial value \cite{chatzipanagiotis2017convergence}. Despite its effectiveness, the primary limitation of MoM is the loss of decomposability, which can hinder its application in large-scale distributed systems. This challenge can be mitigated using a variant of MoM known as the Alternating Direction Method of Multipliers (ADMM). However, for real-time applications, a more versatile framework is required to address diverse scenarios effectively. To this end, we propose employing Graph Neural Networks (GNNs) so as to approximate the solutions given by MoM, thereby achieve possibly better performance compared to conventional optimization methods.
 
\section{Network Routing Optimization using Graph Neural Networks Parameterization} \label{sec:gnn_par}
Note that the optimization problem in \eqref{eq:prob_state} is inherently infinite-dimensional because it requires to determine $a_i^k$ and $r_{ij}^k$ for every possible channel state $\mathbf{R}_t$ and input $a_{0i}^k(t)$. In general, these are challenging to achieve in practical scenarios. As an alternative, we propose a parameterized model that accepts $\mathbf{R}_t$ and $a_{0i}^k(t)$ as inputs and produces the corresponding decisions $a_i^k(t)$, $\mathrm{T}_i^k(t)$ and $\mathrm{K}_{ij}^k(t)$ as outputs. We utilize Graph Neural Networks (GNNs) to parameterize the above routing decisions. Basically, ``GNNs are a specific type of neural network architecture designed to work with graph-structured data'' \cite{battaglia2018relational, wu2020comprehensive}. As mentioned in the recent studies ~\cite{eisen2020optimal, naderializadeh2023learning, naderializadeh2022state, lee2020graph, shen2019graph}, these neural network architectures offer several advantages, including permutation equivariance, scalability, and transferability to different network types.

In the context of our network optimization problem, the network can be denoted by graph $\mathcal{G} = (\mathcal{V}, \mathcal{E}, \mathbf{z}_t, w_t)$ at a particular time instant $t$. Here $\mathcal{V}=[1,2,...,n]$ denotes the set of graph nodes where every node corresponds to a communicating node in the network while $\mathcal{E} \subseteq \mathcal{V} \times \mathcal{V}$ represents the set of directed edges between these nodes. $\mathbf{z}_t$ are the initial node features and $\mathcal{E} \rightarrow \mathbb{R}$ maps each edge to the corresponding weight at a given time $t$, where the weight between a set of 2 interacting nodes is the normalized transmission probability of delivering packets in our optimization problem, i.e. $w_{ij}(t) = R_{ij}(t)$. 

Now considering a single time step $t$ and slightly abusing the notation, once can treat the channel probability matrix $\mathbf{R}_t \in \mathbb{R}^{n \times n}$ to be an adjacency matrix of the network graph, that establishes connections between node $i$ and node $j$. In this context, initial node feature, $\mathbf{z}_t$ serves as the signal associated with the nodes of the network, $i=1,..,n$. Now the GNNs serve as the graph convolutional filter by processing $\mathbf{z}_t \in \mathbf{R}^n$ with the graph structure represented by $\mathbf{R}_t$.
Since we are dealing with a filter, let us denote $\mathbf{h} := \{h_0,...,h_{K-1}\}$ to represent the set of $K$ filter coefficients. The function $\mathbf{\phi}(\mathbf{R}_t)$ is defined as the graph filter, and it is basically a polynomial applied linearly to input signal $\mathbf{z}_t$ on the graph representation \cite{sandryhaila2014big}. The outcome of this convolution process can be expressed as:
\begin{equation} \label{outputy}
    \mathbf{y}_t = \mathbf{\phi}(\mathbf{R}_t)\mathbf{z}_t = \sum_{k=0}^{K-1}h_k \mathbf{R}_t^k \mathbf{z}_t.
\end{equation}
It can be noted observed that the filter $\mathbf{\phi}(\mathbf{R}_t)$ functions as a linear shift-invariant filter, hence the graph $\mathbf{R}_t$ is referred to as Graph Shift Operator (GSO). In a Graph Neural Network (GNN), the input node along with the edge features are processed through $L$ layers where the node features from the previous layer $l-1$, denoted as $\mathbf{Y}_t^{l-1} \in \mathbb{R}^{m \times F_{l-1}}$, are transformed into the node features for the current layer $l$, denoted as $\mathbf{Y}_t^l \in \mathbb{R}^{m \times F_l}$. The filter at layer $l$ is then fed to the output of layer $l-1$ to generate the current layer feature $\mathbf{y}_t^{l}$, that is now expressed as:
\begin{equation} \label{outputyl}
    \mathbf{y}_t^l = \mathbf{\phi}^l(\mathbf{R}_t)\mathbf{z}_t^{l-1} = \sum_{k=0}^{K_l-1}h_{lk} \mathbf{R}_t^k \mathbf{z}_t^{l-1}.
\end{equation}
Subsequently, a pointwise non-linearity $\mathbf{\sigma}$ is then applied to the intermediate features to produce the output of the $l^{th}$ layer:
\begin{equation} \label{outputztl}
    \mathbf{z}_t^l = \mathbf{\sigma}[\mathbf{y}_t^l] = \mathbf{\sigma} \Bigg[ \sum_{k=0}^{K_l-1}h_{lk} \mathbf{R}_t^k \mathbf{z}_t^{l-1} \Bigg].
\end{equation}
Ultimately the Graph Neural Network (GNN) deploys a recursive approach to the convolution operation as shown in \eqref{outputztl}. It is important to note that in \eqref{outputztl}, the non-linearity is applied to each component individually within every layer. Some common choices for the non-linearity function $\mathbf{\sigma}$ comprise of rectified linear units (ReLU), sigmoid functions, or absolute value functions \cite{gama2018convolutional, henaff2015deep}.

While the above equations of GNNs focused on a single graph filter, the GNN's expressive power can be enhanced by implementing a bank of $F_l$ graph filters \cite{eisen2020optimal}. More specifically, this approach allows for the generation of multiple features at each layer, with each feature being processed by a separate filter bank. For instance, the output from layer $l-1$, comprising $F_l$ features, serves as the input for layer $l$ and each of these features is then processed by the $F_l$ filters, $\phi_l^{fg}(\mathbf{R}_t)$. Subsequently, the intermediate feature of the $l^{th}$ layer, after applying these filters, can be represented as
\begin{equation} \label{outputyflg}
    \mathbf{y}_{t,fg}^l = \phi_l^{fg}(\mathbf{R}_t) \mathbf{z}_{t,f}^{l-1} = \sum_{k=0}^{K_l-1}h_{lk}^{fg} \mathbf{R}_t^k \mathbf{z}_{t,f}^{l-1}.
\end{equation}
The expression \eqref{outputyflg} delineates that layer $l$ produces $F_{l-1} \times F_l$ intermediate features $\mathbf{y}_{t,fg}^l$. In order to prevent the potential exponential increase of these features, they are linearly aggregated and processed through the non-linearity function defined by $\mathbf{\sigma}$ to yield output of layer $l$. Therefore, the output, $\mathbf{z}_t^l$ at $l^{th}$ layer is expressed as
\begin{equation}\label{finalzl}
    \mathbf{z}_t^l = \mathbf{\sigma}_l \Bigg[\sum_{f=1}^{F_l} \mathbf{y}_{t,fg}^l \Bigg] = \mathbf{\sigma}_l \Bigg[\sum_{f=1}^{F_l} \phi_l^{fg}(\mathbf{R}_t) \mathbf{z}_{t,f}^{l-1} \Bigg].
\end{equation}
In our experiments, we use \eqref{finalzl} recurrently to generate the GNNs. The filter coefficients are compiled into the filter tensor $\phi = [h_{lk}^{fg}]_{l,k,f,g}$, now the GNN operator can be defined as
\begin{equation}\label{phiout}
   \mathbf{\Psi}(\mathbf{R}_t,\mathbf{x}_t;\phi) = \mathbf{z}_t^L.
\end{equation}
Here $\mathbf{x} = \mathbf{z}_t^0$ is the initial input feature to the GNN network at the first layer, i.e. $l=1$. Depending on our specific algorithms discussed later, we vary the number of input features, $F_0$. Thus, the output layer produces the GNN result, expressed as:
\begin{equation} \label{r_out}
    \mathbf{y_{out}} = \mathbf{Y}_t^L \in \mathbb{R}^{n \times F_L}
\end{equation}
The output from \eqref{r_out} generates a ($n \times F_L$) vector which, after being multiplied to an intermediary matrix $w_r \in \mathbb{R}^{F_L \times F_L}$, helps in determining the packet acceptance decisions $\mathrm{K}_{ij}^k(t)$. To guarantee that the values in the routing decision matrix fall within the probabilistic range of 0 to 1, we process this matrix through a Softmax filter as in
\begin{equation} \label{p_softmax}
    \mathbf{p}(\mathbf{R}_t,\mathbf{x}_t;\bm{\phi}) = \text{Softmax} (\mathbf{y_{out}} ~ w_r ~ \mathbf{y_{out}}^T),
\end{equation}
where $\text{Softmax}(\mathbf{d})_i = \frac{\exp(d_i)}{\sum_{j=1}^{K}\exp(d_j)} $ for $i=1,2,\dots,K$ and $\mathbf{d}=(d_1,d_2,\dots,d_K) \in \mathbb{R}^K$. The intermediate square matrix $w_r$, of dimensions ($F_L \times F_L$), plays a key role in generating the final reception matrix $\mathbf{K}$, which has dimensions ($n \times n$). We then multiply the output of \eqref{r_out} with another intermediary matrix $w_s \in \mathbb{R}^{F_L}$ to obtain the final transmit matrix $\mathbf{T}$. The product matrix is then passed through another Softmax filter to ascertain the values in the probabilistic range of 0 to 1, i.e.
\begin{equation} \label{b_softmax}
   \mathbf{b}(\mathbf{R}_t,\mathbf{x}_t;\bm{\phi}) = \text{Softmax} (\mathbf{y_{out}} ~ w_s).
\end{equation}

In a similar fashion, we determine the auxiliary packets $a_i^k$ by relaying the GNN output via an additional linear layer. This involves multiplying the output to a column vector denoted by $w_a \in \mathbb{R}^{F_L}$, yielding:
\begin{equation} \label{a_ik}
    a_i^k(t) = \big[a_{0i}^k(t) + \mathbf{y_{out}} ~ w_a \big]^+.
\end{equation}
The resultant output $a_i^k(t)$ is further passed through a ReLU filter to ensure the non-negativity of the packets generated at the nodes, reflecting the fact that the number of packet cannot be negative. Note that the GNN is this study differs from our previous work \cite{das2023learning} in that it provides two decision variables for the case of OR isntead of just one. 

\section{Method of Multipliers using Graph Neural Network Parameterization} \label{sec:gnn_mom}
To address the optimization challenge presented in \eqref{eq:prob_state}, which involves the concave objective function, let's apply the Method of Multipliers (MoM). This method aims at finding the optimal solution by leveraging our GNN parameterization as outlined in the previous section. In this setup, we consider $\mathbf{z}_t = a_{0i}^k(t)$ to be the input node feature for our GNN framework, aiming to determine optimal routing policies $\mathbf{p}(\mathbf{R}_t,\mathbf{x}_t;\bm{\phi})$ and $\mathbf{b}(\mathbf{R}_t,\mathbf{x}_t;\bm{\phi})$. By integrating the GNN parameterization expressed in \eqref{phiout}, \eqref{r_out}, \eqref{p_softmax}, \eqref{b_softmax}, we can now reformulate the parameterized network maximization problem initially mentioned in \eqref{eq:prob_state} as:
\begin{subequations} \label{eq:gnn_par}
\begin{align}
    \max_{\phi, w_r, w_s, w_a} \hspace{0.5cm} \mathcal{U} \Biggl(\frac{1}{T} \sum_{t=0}^{T-1} \mathbf{f} \Bigl(\mathbf{R}_t, \mathbf{p}(\mathbf{R}_t;\bm{\phi}), \mathbf{b}(\mathbf{R}_t;\bm{\phi})\Bigr)\Biggr) \\
    s.t. \hspace{1cm} \mathbf{g}\Biggl(\frac{1}{T} \sum_{t=0}^{T-1} \mathbf{f} \Bigl(\mathbf{R}_t, \mathbf{p}(\mathbf{R}_t;\bm{\phi}), \mathbf{b}(\mathbf{R}_t;\bm{\phi}) \Bigr) \Biggr) & \geq 0. \label{gnnconst1}  
\end{align}
\end{subequations}
Accordingly, the Augmented Lagrangian for the network optimization in \eqref{eq:gnn_par} can be expressed as:
\begin{multline} \label{aug_lag1}
    \mathcal{L}(\bm{\phi}, \bm{\mu}) = \mathcal{U} \Biggl(\frac{1}{T} \sum_{t=0}^{T-1} \mathbf{f} \Bigl(\mathbf{R}_t, \mathbf{p}(\mathbf{R}_t;\bm{\phi}), \mathbf{b}(\mathbf{R}_t;\bm{\phi})\Bigr)\Biggr) \\ 
     + \bm{\mu}^T  \mathbf{g}\Biggl(\frac{1}{T} \sum_{t=0}^{T-1} \mathbf{f} \Bigl(\mathbf{R}_t, \mathbf{p}(\mathbf{R}_t;\bm{\phi}), \mathbf{b}(\mathbf{R}_t;\bm{\phi})\Bigr) \Biggr) \\
    + \frac{\rho}{2} \Biggl|\!\Biggl| \mathbf{g}\Biggl(\frac{1}{T} \sum_{t=0}^{T-1} \mathbf{f} \Bigl(\mathbf{R}_t, \mathbf{p}(\mathbf{R}_t;\bm{\phi}), \mathbf{b}(\mathbf{R}_t;\bm{\phi})\Bigr) \Biggl) \Biggr|\!\Biggr|^2
\end{multline}
To facilitate the training of the model parameters $\bm{\phi}_t$, we consider an iterative period $T_0$, representing the number of time steps between successive updates of the model parameters. Abusing the notation for time $t$ slightly, let's consider an iteration index $m \in \{0,1,..., M-1\}$, with $M = \lfloor T/T_0 \rfloor$. The parameters of the model can now be updated in the following manner:
\begin{equation} \label{phim_update_par}
    \bm{\phi}_m = \arg \max_{\bm{\phi} \in \bm{\Phi}} \mathcal{L}(\bm{\phi}, \bm{\mu}_m).
\end{equation}
Subsequently the dual variables, $\bm{\mu}$ can be updated recursively as 
\begin{multline} \label{mum_update_par}
\footnotesize
    \bm{\mu}_{m+1} =  \Bigg[ \bm{\mu}_m \\ - \rho ~ \mathbf{g}\Biggl(\frac{1}{T} \sum_{t=mT_0}^{(m+1)T_0-1} \mathbf{f}(\mathbf{R}_t, \mathbf{p}(\mathbf{R}_t;\bm{\phi}), \mathbf{b}(\mathbf{R}_t;\bm{\phi}) \Bigr) \Biggr) \Bigg]^+.
\end{multline}
Here $\rho$ is the penalty parameter that denotes the learning step size for updating the dual variable. It can be observed that the updates are structured to maximize the Lagrangian $\mathcal{L}$ with respect to $\bm{\phi}$, while ensuring the dual variables $\bm{\mu}$ are adjusted appropriately at each iteration. When run over a sufficiently extended series of time steps, the described routing algorithm is capable of delivering decisions which are both feasible and nearly optimal. It is worth mentioning that conventional dual descent approaches, such as Primal-Dual algorithm, may not always ensure feasibility for the set of routing decisions because the algorithm, as outlined in \eqref{phim_update_par}, \eqref{mum_update_par}, continuously adjusts the routing policy with respect to changes in the dual variables at each iteration. 
However, parameterized policies like this often come with inherent limitations, highlighting the necessity of refining the approach to derive truly optimal routing decisions.

\section{The State Augmentation Algorithm} \label{sec:state_aug}

As highlighted in the previous section, the iterative routing optimization technique mentioned in \eqref{phim_update_par} and \eqref{mum_update_par} is theoretically feasible but encounters several bottlenecks, which significantly impacts the practical implementation of such systems. One major drawback is the need for \textit{a priori} or non-causal knowledge about the state of the network, meaning it requires information about the state of the system at $t=kT_0$. While this might be attainable during training, it is not feasible during testing or execution phase. Another key challenge is the difficulty of achieving convergence to near-optimal system performance, which becomes viable only as the time horizon $T$ approaches infinity. This constraint implies that training of the model parameters cannot conclude after a finite number of iterations. Additionally, there may exist an instant where the iteration index $m$ does not yield a feasible or optimal $\bm{\phi}_m$. Furthermore, the necessity of determining the optimal model parameters in \eqref{phim_update_par} at every time step, given varying set of dual variables $\bm{\mu}_m$, can remarkably increase the computational overhead, especially during the execution stage.

The outlined challenges necessitate the design of an algorithm that eliminates the need for repeatedly retraining the parameters of the network model, $\bm{\phi}_m$ for each new values of $\bm{\mu}_m$. To address this issue, we propose to use the state-augmented routing algorithm, inspired by \cite{calvo2021state, naderializadeh2022state} and our previous work in \cite{das2023learning}. In this method, the state of network at every time instant $t$, represented by $\mathbf{R}_t$, is augmented with the associated dual variables, $\bm{\mu}_{\lfloor t/T_0 \rfloor}$. Subsequently those augmented dual variables, along with the input node features $a_{0i}^k(t)$, are simultaneously provided as inputs to our parameterized GNN model that computes the optimal routing decisions. As a result, the above GNN processes two distinct types of input features, making $F_0=2$. Next, let's consider a unique parameterization to characterize the above state-augmented algorithm, in which expected routing decisions, $\mathbf{p}(\mathbf{R})$ and $\mathbf{b}(\mathbf{R})$ are expressed as $\mathbf{p} (\mathbf{R}, \bm{\mu}; \bm{\theta})$ and $\mathbf{b} (\mathbf{R}, \bm{\mu}; \bm{\theta})$, respectively. Here, the $\bm{\theta} \in \bm{\Theta}$ represents collection of GNN filter tensors that define the state-augmented routing decisions. This process begins with formulation of the augmented Lagrangian, as outlined in \eqref{eq:gnn_par}, for a given set of dual variables $\bm{\mu} \in \mathbb{R}_+^c$, using the expression in \eqref{aug_lag1}:
\begin{multline} \label{eq:aug_lag2}
    \mathcal{L}_{\bm{\mu}}(\bm{\theta}) = \mathcal{U} \Biggl(\frac{1}{T} \sum_{t=0}^{T-1} \mathbf{f} \Bigl(\mathbf{R}_t, \mathbf{p}(\mathbf{R}_t, \bm{\mu};\bm{\theta}), \mathbf{b} (\mathbf{R}_t, \bm{\mu}; \bm{\theta}) \Bigr)\Biggr) \\
    + \bm{\mu}^T  \mathbf{g}\Biggl(\frac{1}{T} \sum_{t=0}^{T-1} \mathbf{f} \Bigl(\mathbf{R}_t, \mathbf{p}(\mathbf{R}_t, \bm{\mu};\bm{\theta}), \mathbf{b} (\mathbf{R}_t, \bm{\mu}; \bm{\theta}) \Bigr) \Biggl) \\
    + \frac{\rho}{2} \Biggl|\!\Biggl| \mathbf{g} \Biggl(\frac{1}{T} \sum_{t=0}^{T-1} \mathbf{f} \Bigl(\mathbf{R}_t, \mathbf{p}(\mathbf{R}_t, \bm{\mu};\bm{\theta}), \mathbf{b} (\mathbf{R}_t, \bm{\mu}; \bm{\theta}) \Bigr) \Biggr) \Biggr|\!\Biggr|^2
\end{multline}
Once we incorporate the specific functional values from the network optimization problem described in \eqref{eq:prob_state}, the augmented Lagrangian takes the following form,
\begin{multline} \label{eq:lagrangian_applied}
\scriptsize
    \mathcal{L}_{\bm{\mu}} (\bm{\theta}) = \sum_k \sum_i \log \Biggl(\frac{1}{T} \sum_{t=0}^{T-1} a_i^k(t) \Biggl)
    + \bm{\mu}_i^k \Bigg[ \frac{1}{T} \\ \sum_{t=0}^{T-1} \bigg( \mathrm{T}_i^k(t) \mathrm{C}_i(t) 
    - a_i^k(t) - \sum_{j \in n_i} \mathrm{T}_j^k(t) \mathrm{R}_{ij}(t) \mathrm{K}_{ij}^k(t)a_j^k(t) \\ - z_i^k(t) \bigg)\Bigg] 
    - \frac{\rho}{2} \Biggl|\!\Biggl| \mathrm{T}_i^k(t) \mathrm{C}_i(t) - a_i^k(t) \\ - \sum_{j \in n_i} \mathrm{T}_j^k(t) \mathrm{R}_{ij}(t) \mathrm{K}_{ij}^k(t)a_j^k(t) - z_i^k(t) \Biggr|\!\Biggr|^2.
\end{multline}
Here the Lagrangian multipliers $\bm{\mu}_i^k$ correspond to the flow constraint (\ref{eq:7b}) while and the constraints \eqref{eq:7c} and \eqref{eq:7c} are implicitly maintained.
We then consider the dual variables drawn from a probability distribution $p_{\bm{\mu}}$. The state-augmented routing policy can now be defined as the one that maximizes the expected value of the augmented Lagrangian over the probability distribution of all parameters, i.e.
\begin{equation} \label{eq:theta*_sa}
    \bm{\theta}^* = \arg \max_{\bm{\theta} \in \bm{\Theta}} \mathbb{E}_{\bm{\mu} \sim p_{\bm{\mu}}} \big[\mathcal{L}_{\bm{\mu}} (\bm{\theta}) \big]
\end{equation}
This state-augmented policy, which is now parameterized by $\bm{\theta}^*$, facilitates the search for optimal Lagrangian-maximized routing decisions $\mathbf{p}(\mathbf{R}, \bm{\mu};\bm{\theta})$ and $\mathbf{b}(\mathbf{R}, \bm{\mu};\bm{\theta})$ for each iteration of the dual variable $\bm{\mu} = \bm{\mu}_m$. Utilizing this concept, the update for the dual variable at iteration $m$ in \eqref{eq:theta*_sa} can be expressed as,
\begin{multline} \label{eq:mu_sa1}
\small
    \bm{\mu}_{m+1} = \Bigg[ \bm{\mu}_m - \gamma_{\bm{\mu}} \times \\
    \mathbf{g}\Biggl(\frac{1}{T_0} \sum_{t=mT_0}^{(m+1)T_0-1} \mathbf{f} \Bigl(\mathbf{R}_t, \mathbf{p}(\mathbf{R}_t, \bm{\mu}_m;\bm{\theta}^*), \mathbf{b}(\mathbf{R}_t, \bm{\mu}_m;\bm{\theta}^*) \Bigr) \Biggl) \Bigg]^+
\end{multline}
During the execution phase, the update for the dual variable, based on the network optimization problem in \eqref{eq:prob_state}, is calculated as,
\begin{multline} \label{eq:mu_sa2}
    \bigg[\bm{\mu}_i^k(t)\bigg]_{m+1} = \bigg[\bm{\mu}_i^k(t)\bigg]_m 
    - \gamma_{\bm{\mu}} \Bigg[ \frac{1}{T} \sum_{t=0}^{T-1} \Bigg( \mathrm{T}_i^k(t) \mathrm{C}_i(t) \\ 
    - a_i^k(t) - \sum_{j \in n_i} \mathrm{T}_j^k(t) \mathrm{R}_{ij}(t) \mathrm{K}_{ij}^k(t)a_j^k(t) \bigg) \Bigg]
\end{multline}

The equation above facilitates simultaneous learning of a parameterized model and the derivation of the optimal state-augmented routing policy, as outlined in \eqref{eq:theta*_sa}. This process effectively mirrors the gradient descent steps characteristic of traditional dual descent algorithms. Given that the resulting routing matrices are derived from localized information processed on channel probability matrices, this operation is best interpreted within the framework of graph signal processing.

\begin{subsection}{Implementing State-Augmentation and Practical Assumptions}

\begin{algorithm*}
\caption{State-Augmented Routing Optimization: Training Phase}
\label{alg:alg1}
\begin{algorithmic}[1]
 \renewcommand{\algorithmicrequire}{\textbf{Input:}}
 \REQUIRE T, number of time steps, number of training iterations $N_{train}$, batch size B,  primal learning rate $\gamma_\theta$
\STATE \textit{Initialization} : $\bm{\theta}_0, q_i^k(0)$ 
\FOR{$n=0,...,N_{train} -1$}
\FOR{$b=0,...,B-1$}
\STATE dual variables are randomly sampled  ~$\bm{\mu}_b \sim p_{\mu}$
\STATE A sequence of channel states is randomly generated  $\{\mathbf{R}_{b,t} \}_{t=0}^{T-1}$
\FOR{$t=0,1,...,T-1$}
\STATE Routing decisions obtained: $\mathbf{p}(\mathbf{R}_{b,t},\mathbf{x}_b;\bm{\theta}_n))$ and $\mathbf{b}(\mathbf{R}_{b,t},\mathbf{x}_b;\bm{\theta}_n))$
\STATE Queue length are evaluate by \eqref{queueupdate}
\ENDFOR
\STATE The augmented Lagrangian is evaluated from \eqref{eq:aug_lag2}, i.e.
\begin{multline*}
    \mathcal{L}_{\bm{\mu}_b}(\bm{\theta}) = \mathcal{U} \Biggl(\frac{1}{T} \sum_{t=0}^{T-1} \mathbf{f} \Bigl(\mathbf{R}_{b,t}, \mathbf{p}(\mathbf{R}_{b,t}, \mathbf{x}_b;\bm{\theta}_n), \mathbf{b}(\mathbf{R}_{b,t},\mathbf{x}_b;\bm{\theta}_n) \Bigr)\Biggl) \\ 
    + \bm{\mu}^T  \mathbf{g}\Biggl(\frac{1}{T} \sum_{t=0}^{T-1} \mathbf{f} \Bigl(\mathbf{R}_{b,t}, \mathbf{p}(\mathbf{R}_{b,t}, \mathbf{x}_b;\bm{\theta}_n),\mathbf{b}(\mathbf{R}_{b,t},\mathbf{x}_b;\bm{\theta}_n) \Bigr) \Biggl) 
    + \frac{\rho}{2} \Biggl|\!\Biggl| \mathbf{g}\Biggl(\frac{1}{T} \sum_{t=0}^{T-1} \mathbf{f} \Bigl(\mathbf{R}_{b,t}, \mathbf{p}(\mathbf{R}_{b,t}, \mathbf{x}_b;\bm{\theta}_n), \mathbf{b}(\mathbf{R}_{b,t},\mathbf{x}_b;\bm{\theta}_n) \Bigr) \Biggl) \Biggr|\!\Biggr|^2
\end{multline*}
\ENDFOR
\STATE The model parameters , $\bm{\theta}$, are updated by (\ref{eq:par_update_sa})
\begin{equation*}
    \bm{\theta}_{n+1} = \bm{\theta}_{n} + \frac{\gamma_\theta}{B} \sum_{b=1}^{B-1} \nabla_\theta \mathcal{L}_{\bm{\mu}_b} (\bm{\theta}_n)
\end{equation*}
\ENDFOR
\STATE $\bm{\theta}^* \leftarrow \bm{\theta}_{N_{train}}$
\renewcommand{\algorithmicrequire}{\textbf{Output:}}
 \REQUIRE Save the optimal parameters of the model $\theta^*$
\end{algorithmic}
\end{algorithm*}

The maximization process described in \eqref{eq:theta*_sa} is required to be performed during the offline training stage. In this phase, we leverage a regular gradient ascent method to identify and learn an optimal set of model parameters, i.e. $\bm{\theta}^*$. Once the training phase is completed, these parameters are generally saved for use in the execution phase. Throughout the model training stage, a batch of dual variables, denoted as $\{ \bm{\mu}_b \}_{b=1}^{B}$, is processed. They are randomly drawn from the probability distribution $p_{\bm{\mu}}$. Therefore, the practical mode of achieving the Lagrangian maximization specified in \eqref{eq:theta*_sa} can be expressed in its empirical form as below.
\begin{equation} \label{eq:16}
    \bm{\theta}^* = \arg \max_{\bm{\theta} \in \bm{\Theta}} \frac{1}{B} \sum_{b=1}^{B-1} \mathcal{L}_{\bm{\mu}_b} (\bm{\theta})
\end{equation}
Similar to optimizing the primal variable in a primal-dual learning, the optimization problem in consideration can be solved iteratively via the method of gradient ascent. We begin this process by a random initialization of the model parameters, denoted as $\theta_0$, which are then progressively updated throughout the iterations, indexed as $n=0,...,N_{train}-1$. The formula for updating the parameter $\bm{\theta}$ is given as,
\begin{equation} \label{eq:par_update_sa}
    \bm{\theta}_{n+1} = \bm{\theta}_{n} + \frac{\gamma_\theta}{B} \sum_{b=1}^{B-1} \nabla_\theta \mathcal{L}_{\bm{\mu}_b} (\bm{\theta}_n),
\end{equation}
where $\gamma_\theta$ denotes the learning rate or step size for the model parameters $\theta$. It is worth noting that these parameters correspond to the coefficients of graph filters within our defined GNN framework. Once the model parameters are optimized using \eqref{eq:par_update_sa}, the GNN follows to update the primal variables $\{\bm{a},\mathbf{T, K}\}$. After the model's training phase is complete, we save the resulting set of optimized parameters which have reached convergence, as $\bm{\theta}^*$. In order to enhance the model's capacity to generalize to random network states, we couple each set of dual variables, $\{\bm{\mu}_b\}_{b=0}^{B-1}$ with a uniquely sampled sequence of network states $\{\mathbf{R}_{b,t} \}_{t=0}^{T-1}$. This approach significantly helps in optimizing the model across a family of network scenarios including time varying conditions. The detailed steps of this training methodology are comprehensively presented in Algorithm \ref{alg:alg1}.

Upon the completion of the training stage, we proceed to the execution phase, during which once can update the dual variables iteratively to compute the routing decisions based on the current channel state. 
At the start of the execution phase, we set the dual variables to zero, $\bm{\theta} = \bm{\theta}_0$. For each time instant $\{t\}_{t=0}^{T-1}$, and for a specific network state $\bm{R}_t$, the optimal network decisions are delivered by above state-augmented policies $\mathbf{p}(\bm{R}_t,\mathbf{x}_{\lfloor t/T_0 \rfloor};\bm{\theta}^*)$, as determined during Algorithm \ref{alg:alg1}. Next, the dual multipliers follow an update at every $T_0$ time step in accordance with \eqref{eq:mu_sa2}. We can observe that the dynamics of the dual variables in the execution phase, as expressed in \eqref{eq:mu_sa2}, can facilitate the amenable satisfaction of the constraints. As detailed in Algorithm 2, optimal routing decisions generated at any given instant $t$ leads to constraint satisfaction if the dual variables get minimized. Conversely, an increase of the dual variable values indicates a constraint violation, signaling the need for fine-tuning certain algorithm parameters during the execution phase. The methodologies illustrated in both algorithms are largely in line with those in \cite{naderializadeh2022state}, although with minor modifications designed to suit our specific problem setup.

\begin{algorithm*}
\caption{State-Augmented Routing Optimization: Execution Phase.}
\label{alg:alg2}
\begin{algorithmic}[1]
 \renewcommand{\algorithmicrequire}{\textbf{Input:}}
 \REQUIRE Optimal model parameters $\theta^*$, iteration time $T_0$, sequence of network states $\{\mathbf{R}_{b,t} \}_{t=0}^{T-1}$, dual learning rate $\gamma_\mu$
\STATE \textit{Initialization} : $\bm{\mu}_0 \leftarrow 0, m \leftarrow 0$
\FOR{$t=0,1,...,T-1$}
\STATE Routing decisions are generated: $\mathbf{p}_t$ := $\mathbf{p}(\mathbf{R}_t,\mathbf{x}_m;\bm{\theta}^*))$ and $\mathbf{b}_t$ := $\mathbf{b}(\mathbf{R}_t,\mathbf{x}_m;\bm{\theta}^*))$
\STATE queue length is evaluated using (\ref{queueupdate})
\IF{$(t+1)$ mod $T_0$ = 0}
\STATE The dual variables, $\bm{\mu}$, updated by (\ref{eq:mu_sa1})
\begin{equation*} 
    \bm{\mu}_{m+1} = \Bigg[ \bm{\mu}_m 
    - \gamma_{\bm{\mu}} \mathbf{g}\Biggl(\frac{1}{T_0} \sum_{t=mT_0}^{(m+1)T_0-1} \mathbf{f} \Bigl(\mathbf{R}_t, \mathbf{p}(\mathbf{R}_t,\mathbf{x}_m;\bm{\theta}^*) , \mathbf{b}(\mathbf{R}_t,\mathbf{x}_m;\bm{\theta}^*) \Bigr) \Biggl) \Bigg]^+
\end{equation*}
$m \leftarrow m+1$
\ENDIF
\ENDFOR
\renewcommand{\algorithmicrequire}{\textbf{Output:}}
 \REQUIRE Obtain the sequence of network routing decisions $\{\mathbf{p}_t, \mathbf{b}_t\}_{t=0}^{T-1}$
\end{algorithmic}
\end{algorithm*}
\end{subsection}

\section{Numerical Results \\ \& Observations} \label{sec:results}
\subsection{Generating the Network Architectures}
We begin our experimental process by creating random geometric network graphs consisting of $N=|\mathcal{V}|$ nodes. To generate these networks, we use the $k$-Nearest Neighbor ($k$-NN) method and place $N$ communicating wireless nodes within a unit circle. We set $k=4$ for our simulations involving random graphs and the transmission capacity of each node $\mathrm{C}_i$ is set to 100. The training model architecture utilizes a 3-layer Graph Neural Network (GNN), with the layers configuration set to posses $F_0 = 2, F_1=16$, and $F_2=8$ features, respectively. For training the GNN model, we leverage the ADAM optimizer while setting a primal step size of $\gamma_{\bm{\theta}} = 0.05$ for optimization of the primal model parameters. Additionally, we set the penalty term $\rho = 0.005$, while it undergoes exponential decay to assist in optimization of the dual variable. We conduct the simulations over $T=100$ time steps, with parameter updates being carried out every $T_0=5$ steps. For simplicity, we calculate the channel state probability $\mathrm{R}_{ij}$ between two nodes using a linear function,
\begin{equation} \label{eq:channel_prob}
    \mathrm{R}_{ij} = 1 - \frac{d_{ij}}{d_c},
\end{equation}
where $d_{ij}$ is the distance between a pair of nodes, node $i$ and node $j$ while $d_c$ represents the cutoff distance after which delivery of packets along the link drops to zero. Our model captures a declining trend in the link performance with increase in distance similar to more sophisticated probabilistic channel models like the ones used in ~\cite{mox2020mobile}

The analysis of the state-augmented model parameters in the previous section consider the use of time-varying channel probabilities whereas we use a constant channel according to \eqref{eq:channel_prob}. Therefore, the objective now is to determine the policy $\mathbf{p}(\mathbf{R},\mathbf{x};\bm{\theta}^*)$ for $\mathbf{R} \in \mathcal{R}$. For a network of given size, we generate 128 training samples and 16 testing samples, using a batch size of 16 samples in each batch. We sample the dual variables randomly from a uniform distribution $U(1,5)$ for the training phase and run the training iterations for 30 epochs. The implementation code can be accessed at \url{https://github.com/sourajitdas/State-Augmented-Routing-Wireless-Communication.git}.

\subsection{Performance Across Different Unparameterized Algorithms}
\begin{figure}[h]
    \centering   
    \includegraphics[width=3.5in]{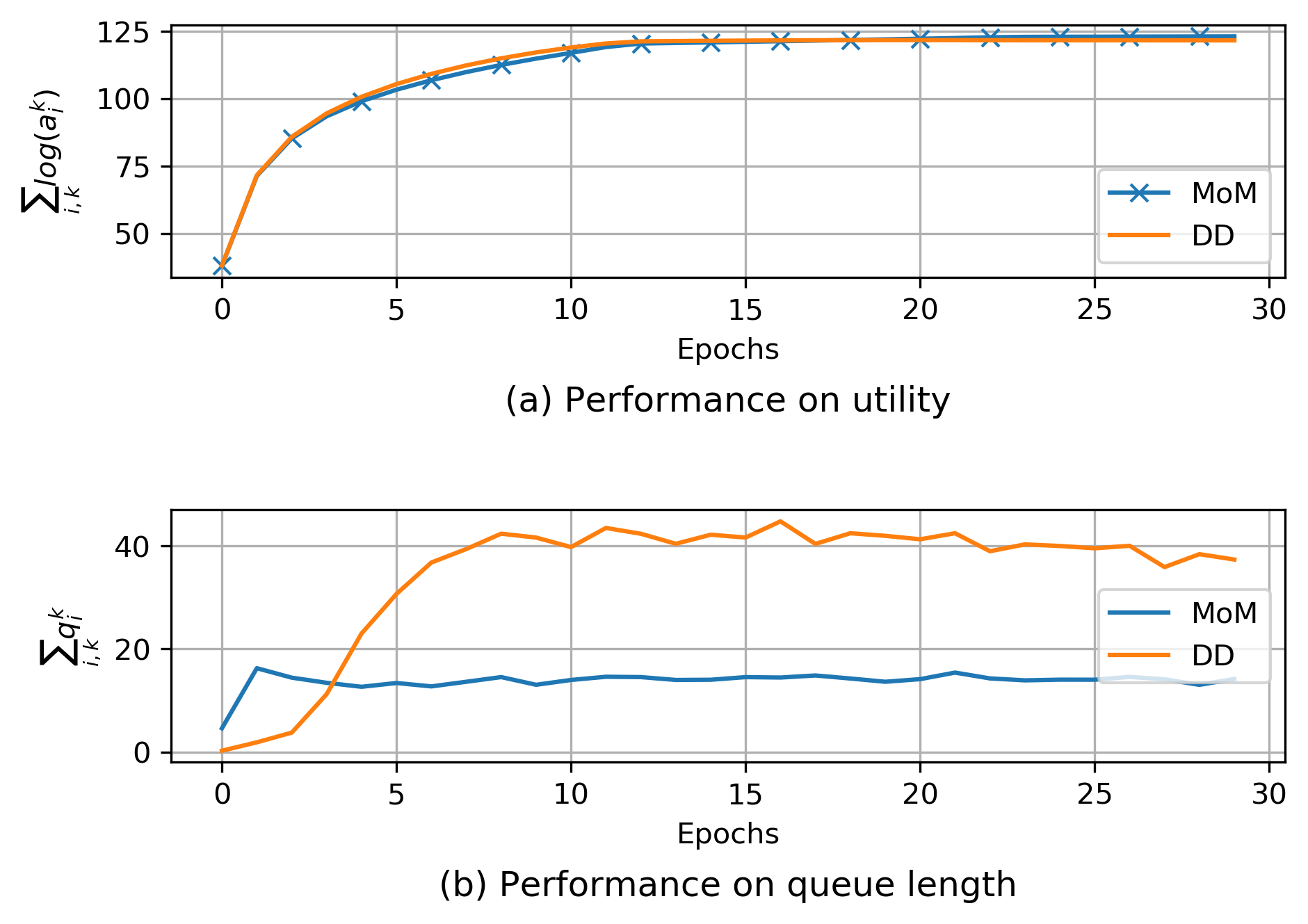}
    \caption{Performance comparison between two unparameterized approaches—Method of Multipliers (MoM) and Dual Descent (DD)—for a network configuration with 10 nodes and 4 flows, evaluated over a single time step.}
    \label{Fig:unpar_algo}
\end{figure}


We begin by evaluating the performance of two unparameterized optimization methods on networks consisting of $N=10$ nodes and $K=4$ flows. To limit the total number of optimization variables, all methods are executed for 30 epochs with $T=1$. Fig.~\ref{Fig:unpar_algo} illustrates the performance of the two unparameterized approaches: Method of Multipliers (MoM) and Dual Descent (DD). From the utility plot, it is evident that DD demonstrates slower convergence toward the optimal solution, whereas the augmented Lagrangian method (MoM) achieves better results because it maximizes the utility function while reducing the queue lengths. This behavior is also reflected in the queue length performance, as slower convergence in DD leads to an accumulation of more packets in the node queues.

\subsection{Proposed State-Augmentation vs Other Approaches}
\begin{figure}[h]
    \centering   
    \includegraphics[width=3.5in]{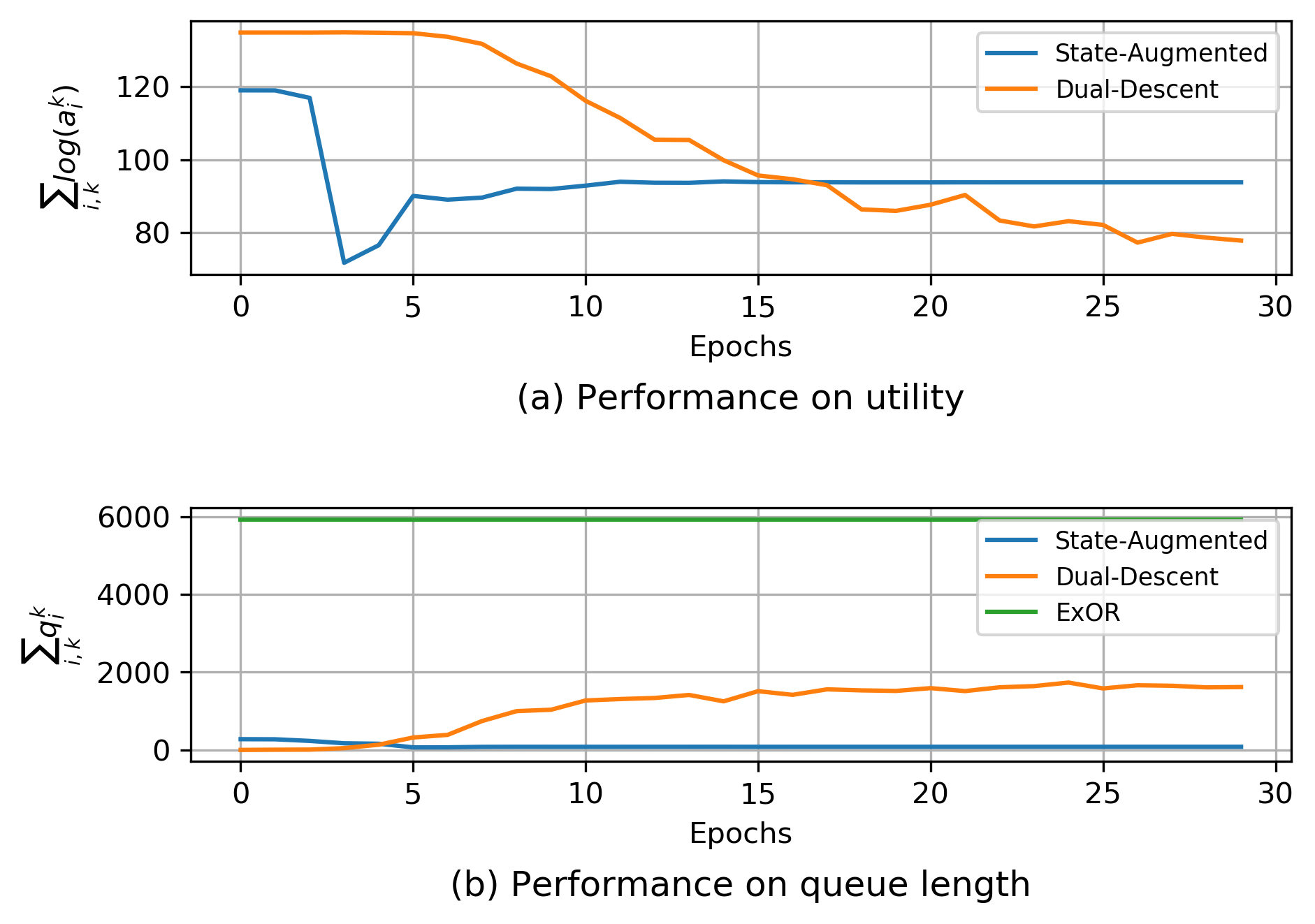}
    \caption{Comparison of performance between the parameterized state-augmented approach utilizing GNNs, the unparameterized Dual Descent method and ExOR protocol for networks consisting of 10 nodes and 4 flows, evaluated over $T=100$ time steps.}
    \label{Fig:par_vs_unpar}
\end{figure}
Next, we evaluate the trained state-augmented GNN model on randomly generated networks consisting of 10 nodes and 4 flows for $T = 100$ time intervals. The performance of the GNN model is compared against the primal-dual algorithm based on the dual-descent method and the Extremely Opportunistic Routing (ExOR) protocol as baseline \cite{biswas2005ExOR}. As illustrated in Fig~\ref{Fig:par_vs_unpar}, our model demonstrates superior performance compared to the traditional dual descent approach. Not only does our model nearly match the optimal utility achieved by primal-dual as shown in Fig.~\ref{Fig:par_vs_unpar}(a), it also surpasses primal-dual and ExOR protocol in reducing queue lengths, as seen in Fig. \ref{Fig:par_vs_unpar}(b). The ExOR protocol being a conventional non-learning routing protocol fails to catch up with the increasing packet arrivals as the time progresses.

\begin{figure}[h]
    \centering   
    \includegraphics[width=3.5in]{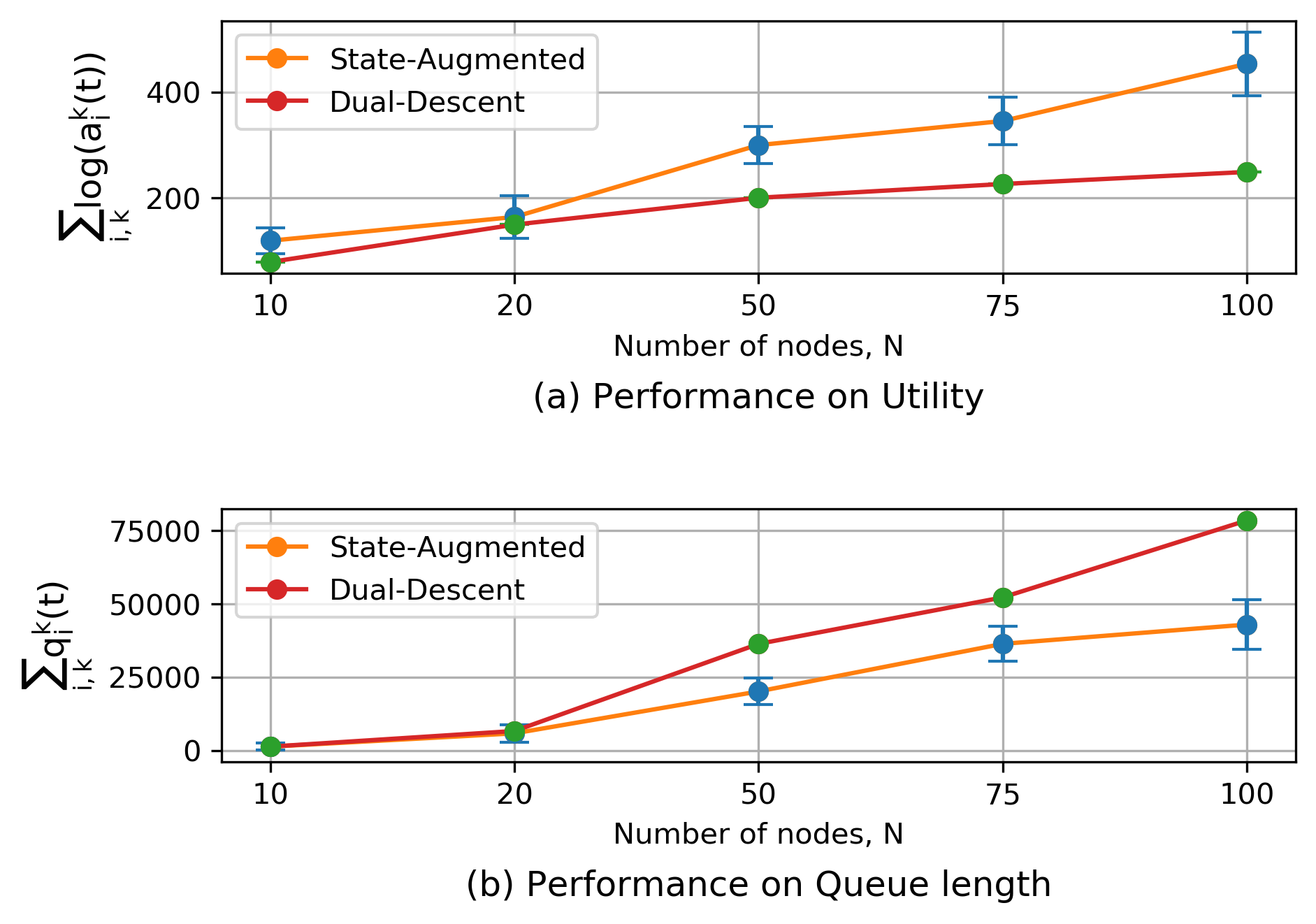}
    \caption{Comparison of the performance of state-augmented and dual descent algorithms for networks with 4 flows and varying node counts ($N\in\{10, 20, 50, 75, 100\}$).}
    \label{Fig:sa_vs_dd_nodes}
\end{figure}
Next we increase the network size from 10 to 100 nodes and observe both the State Augmentation (SA) model and primal-dual learning show improved utility performance. Despite the increased variance observed with the parameterized SA model, its average performance remains comparably close to that of primal-dual method, as shown in Fig.~\ref{Fig:sa_vs_dd_nodes}a. A similar trend is observed in the queue length stability for both methods, as depicted in Fig.~\ref{Fig:sa_vs_dd_nodes}b. 

\begin{figure}[h]
    \centering   
    \includegraphics[width=3.5in]{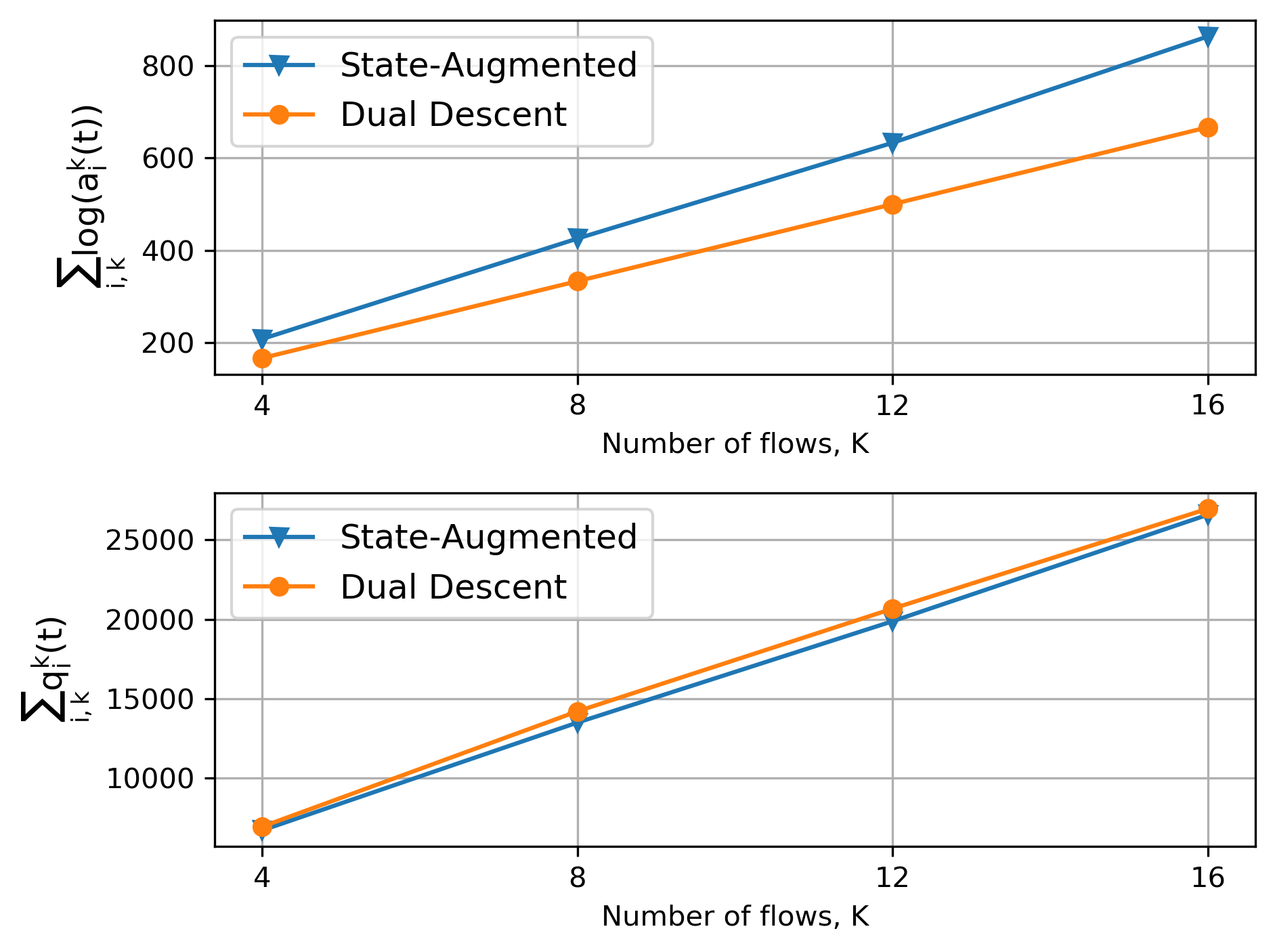}
    \caption{Comparison of state-augmented and dual descent algorithms for a network with 50 nodes and varying flow counts ($K\in\{4, 8, 12, 16\}$).}
    \label{Fig:sa_vs_dd_flows}
\end{figure}
In another simulation, with the network size fixed at 50 nodes while increasing the number of flows from 4 to 16, we observe that both algorithms exhibit improved performance as the number of flows increases, with state augmentation maintaining performance levels close enough to the optimal results of primal-dual learning (Fig~\ref{Fig:sa_vs_dd_flows}(a)). The increase in the number of flows also results in larger queue sizes since increase in the flow numbers lead to increased accumulation of packets in the network. The above results showcase the acclaimed \emph{scalability} properties of GNN-based solutions compared to traditional methods.

\begin{figure}[h]
    \centering   
    \includegraphics[width=3.5in]{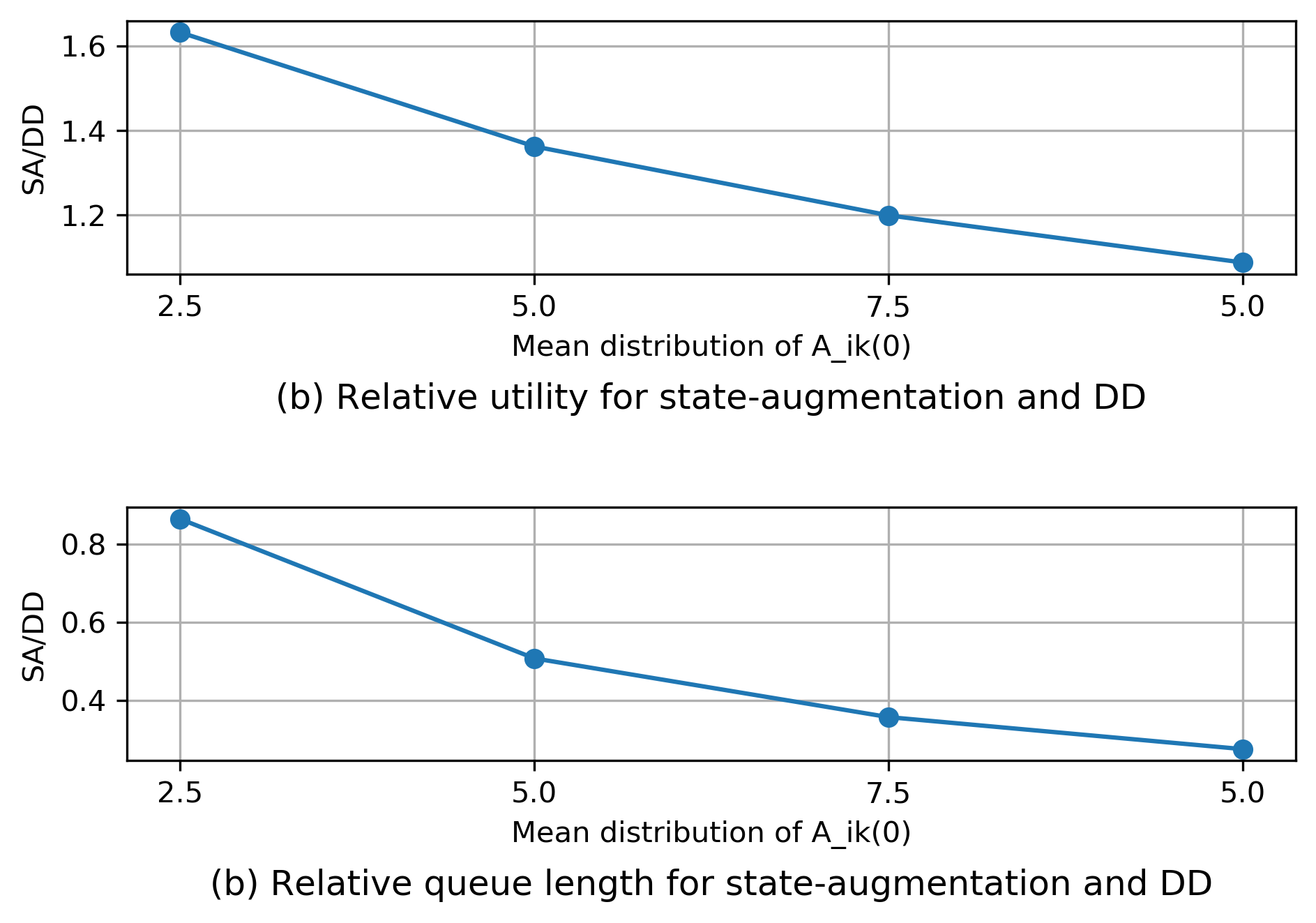}
    \caption{Performance of state-augmented algorithm relative to the dual descent for a random network with 50 nodes and 4 flows.}
    \label{Fig:rel_comp}
\end{figure}

We further analyze the comparison of the state augmentation algorithm's relative performance to dual descent algorithm in a network of 50 nodes and 4 flows. By varying the average of the input to the GNN, which is  $a_{0i}^k(t)$, across five different traffic conditions, we observe that the relative performance of state augmentation to dual descent decreases with increase in the traffic input, as shown in Fig.~\ref{Fig:rel_comp}(a). This shows that the state-augmented GNN model has a reduced efficiency while handling increased traffic, thereby increasing the queue length stability, a fact that is supported by the plot in Fig.~\ref{Fig:rel_comp}(b). 
\subsection{Stability to Perturbation}
Another interesting characteristics of GNN to consider is the \emph{stability} property. We considered a random dataset and introduced perturbation to 50\% of the nodes within a network, shifting their positions by 20\% from the original ones. Note that such an adjustment led to a unanimous increase or decrease in the network edges, resulting in a newly structured graph for each sample. 
\begin{figure}[h]
    \centering   
    \includegraphics[width=3.5in]{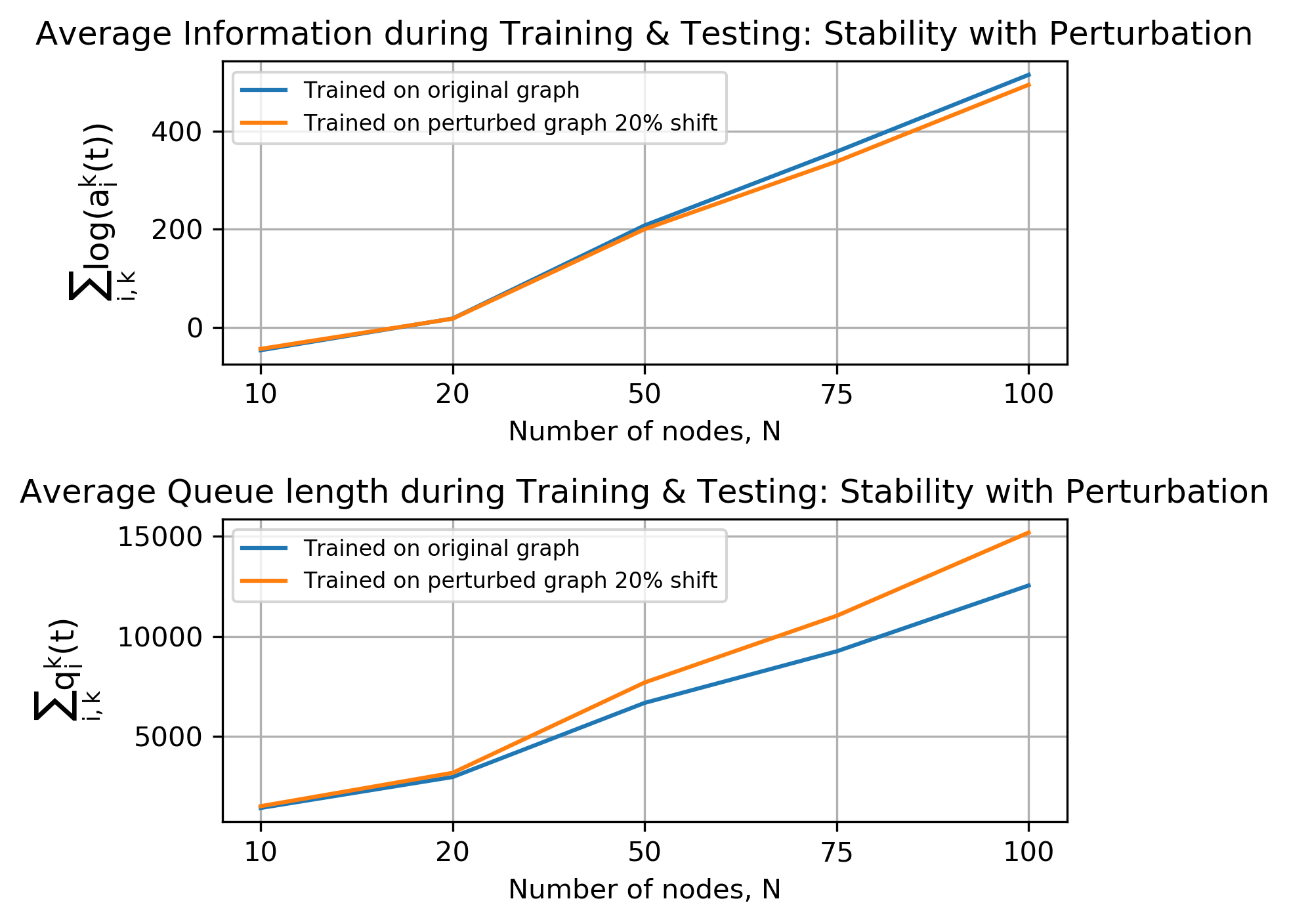}
    \caption{Evaluation of the proposed state augmentation based GNN with respect to perturbation of network nodes.}
    \label{Fig:perturb}
\end{figure}
As observed in Fig.\ref{Fig:perturb}, the GNN model delivers commendable performance on these perturbed graphs with respect to the original ones, affirming its stability property. However, it is crucial to acknowledge that slight variances may occur as the node count may change due to the significant variation in the number of edge modifications. This scenario can be potentially applicable in dynamic network environments, such as in multi-agent systems where nodes are mobile and frequently change their locations. This demonstrates the GNN's extensive ability to adapt to changing network structures while maintaining its performance even though the spatial configuration of the nodes varies.

\subsection{Transference to Unknown Networks of different sizes}
As discussed in Section~\ref{sec:gnn_par}, one of the key features of GNNs is their ability to generalize across different network sizes, a property often referred to as size invariance. This characteristic enables GNNs to operate effectively on networks that were not encountered during training. To evaluate this, a state-augmented GNN model was trained on a network with $N = 20$ nodes and $K = 5$ flows. The trained model was then tested on networks with node counts varying from 10 to 100. 
\begin{figure}[h]
    \centering   
    \includegraphics[width=3.5in]{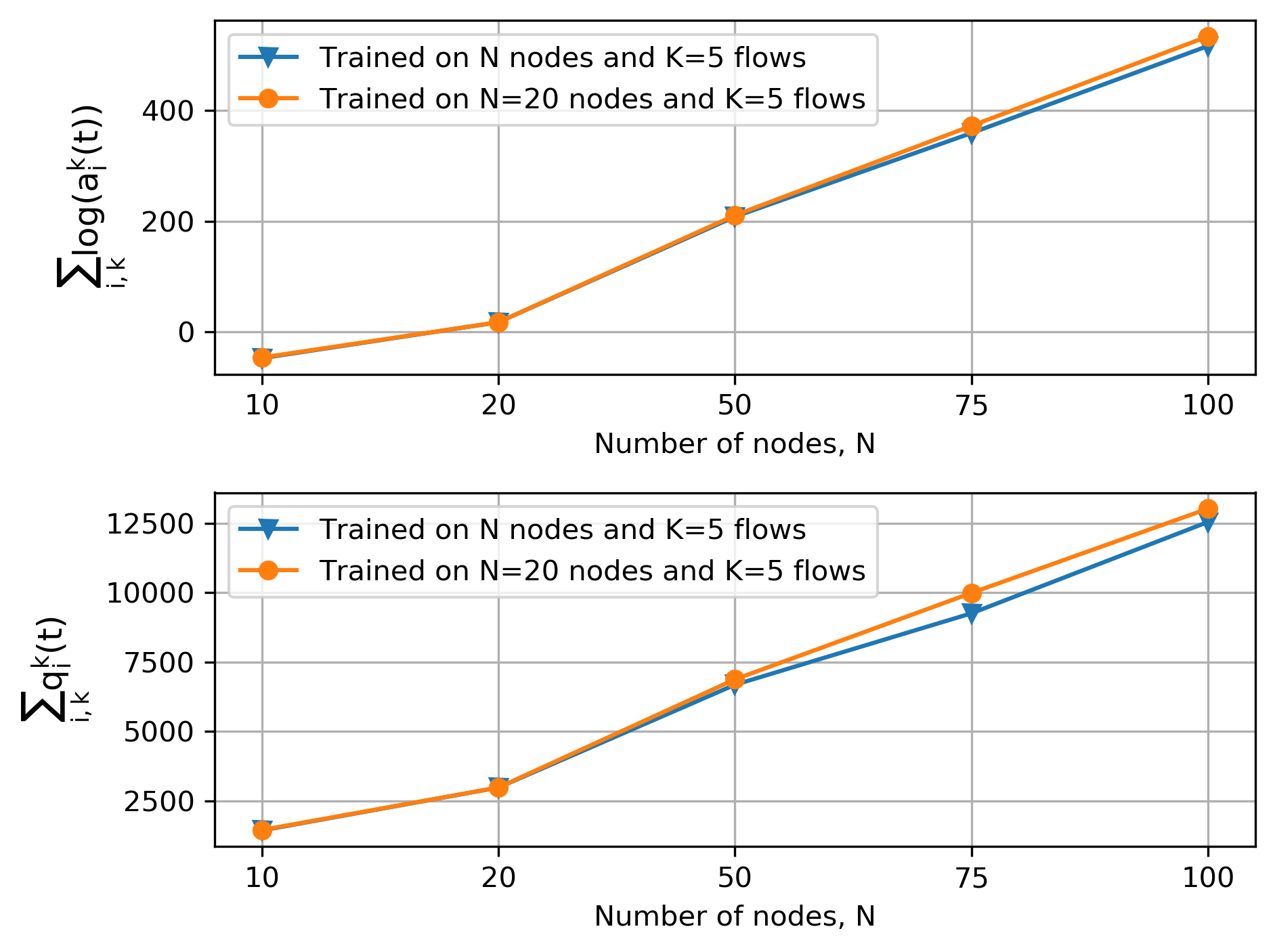}
    \caption{Evaluation of the transference of the proposed state-augmented algorithm on networks of varying sizes, trained on a network with 20 nodes and 5 flows.}
    \label{Fig:transfer_nodes}
\end{figure}
As seen in Fig.~\ref{Fig:transfer_nodes}, it illustrates a 20 node network trained GNN model's \emph{transference} or \emph{transferability} properties to adapt to networks of previously unseen sizes. We observe superior performance of the already trained model particularly on larger graphs, although there is a slight increase in queue length for network with higher number of nodes. Additionally, tests conducted on a 20-node network demonstrate the GNN's robust and consistent performance across different random graphs of the same size.

We continue with similar tests where a GNN model was trained on a network with $N = 50$ nodes and $K = 10$ flows and then evaluated on a 50-node network with varying numbers of flows. 
\begin{figure}[h]
    \centering   
    \includegraphics[width=3.5in]{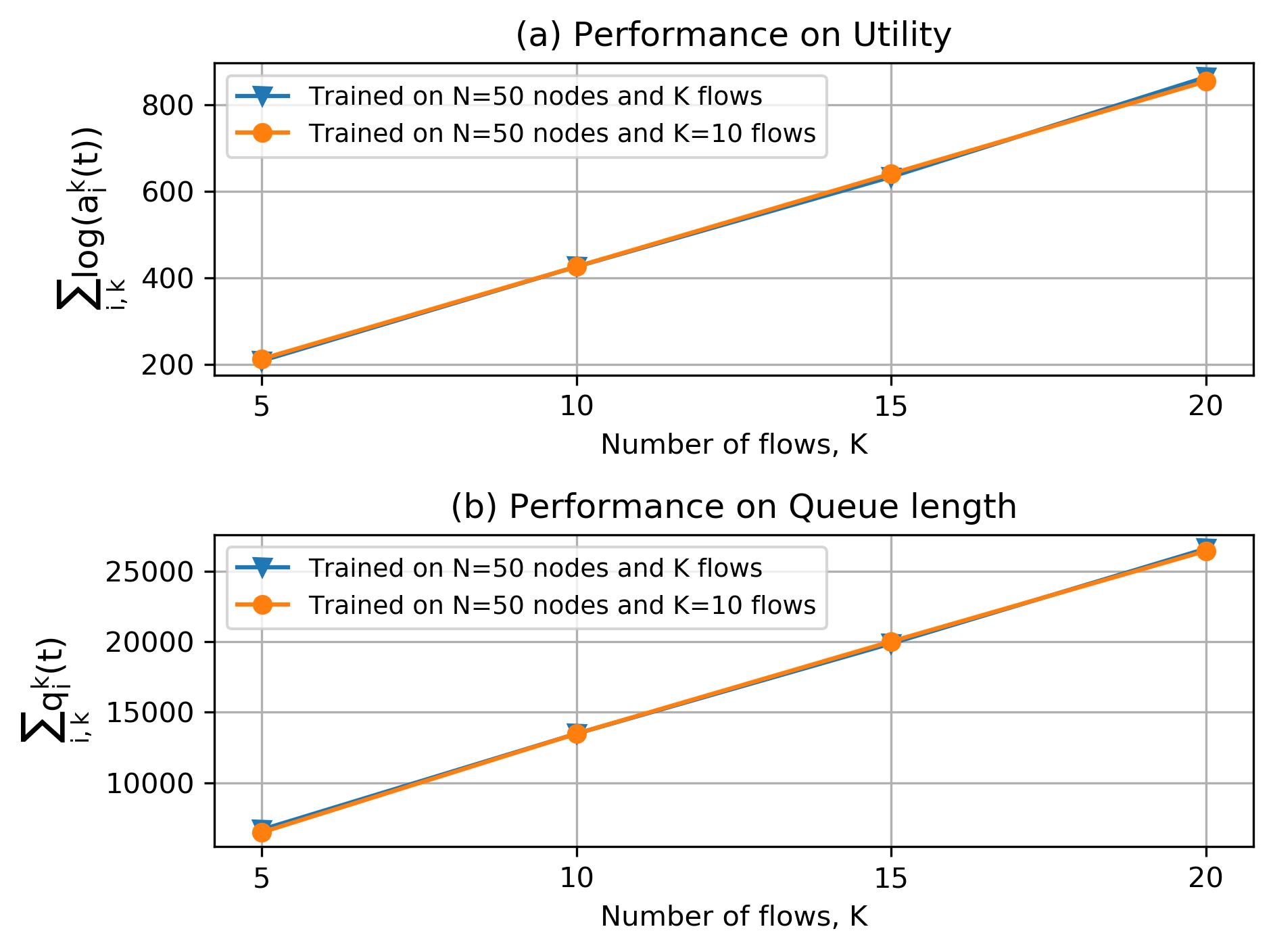}
    \caption{Evaluation of the transference of the proposed state-augmented algorithm to networks with varying flow counts, trained on a network with 50 nodes and 10 flows.}
    \label{Fig:transfer_flows}
\end{figure}
As shown in Fig.\ref{Fig:transfer_flows}, the model successfully maintained its performance across different flow configurations, thereby proving its robust transferability property. The network utility and queue length stability of the trained model were comparable to those achieved by the models which were initially trained on the same graph. Furthermore, scenario with 10 flows reinforces the GNN's transferability to networks of the same size and flow count. This property of GNNs is particularly beneficial in real time applications, as it enables the training of models on smaller networks offline and their subsequent execution on larger networks online, thereby offering significant savings in computational resources.

\subsection{Route formation and packet handling}
\begin{figure}[h]
    \centering   
    \includegraphics[width=3.5in]{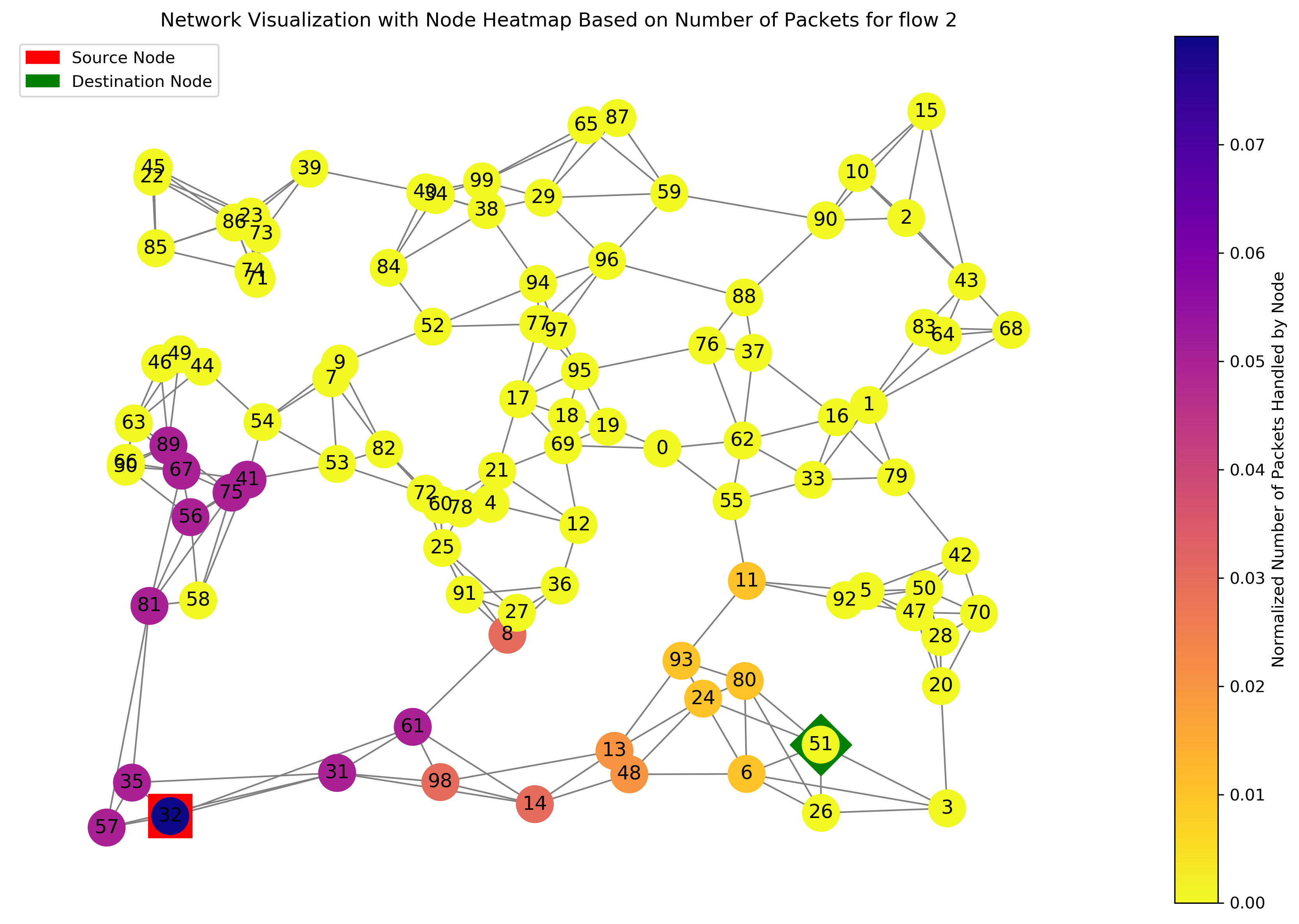}
    \caption{Evolution of the route on a network with 100 nodes and 4 flows showing the transfer of packets from source to destination for a particular flow.}
    \label{Fig:route_map}
\end{figure}
An interesting observation is to find out the paths traced by the model to transmit the information packets from source node $i$ to the destination node $o_k$. We run the trained state-augmented GNN model on a sample test network with 100 nodes and 4 flows. As shown in Fig.\ref{Fig:route_map}, we plot the normalized packet handled in the network for a particular flow at each node. It is clearly evident that the GNN does a spectacular job in routing the packets from the source node through to the final destination node since majority of packets are handled by the nodes that lie in between the source and node.

\subsection{Behavior of Dual variable and Stability of Queue length}
\begin{figure}[htp]
    \centering   
    \includegraphics[width=3.5in]{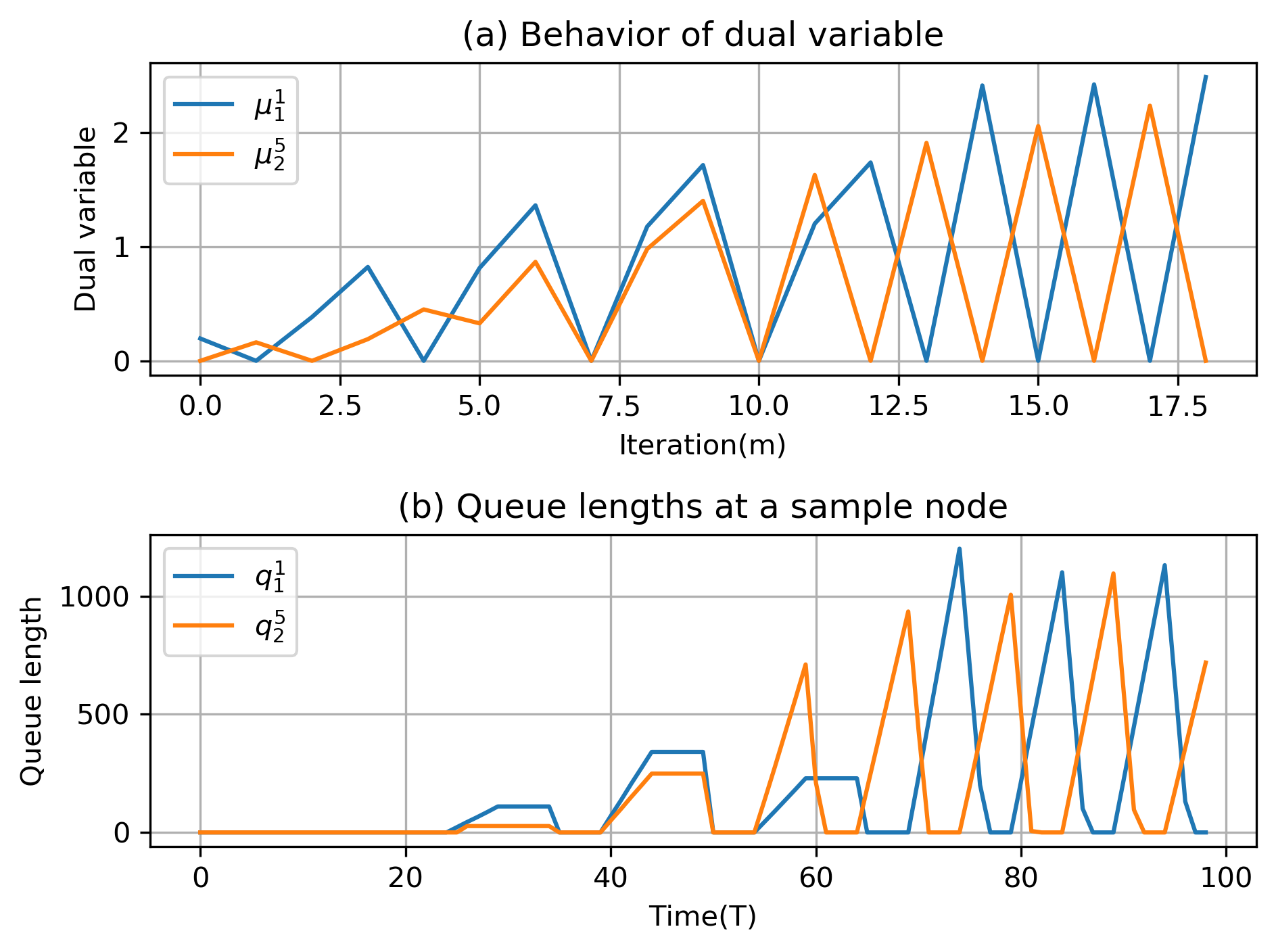}
    \caption{Dual variable behavior and stability of queue length for a sample node in a network with 20 nodes and 4 flows.}
    \label{Fig:dual_perform}
\end{figure}
We analyze network samples with $N = 20$ nodes and $K=4$ flows, training the model over $T = 100$ time steps to investigate the behavior of dual variables and their influence on network performance. Fig. \ref{Fig:dual_perform} provides a graphical representation of the queue lengths at each time step $t$ alongside the associated dual variables $\bm{\mu}_{\lfloor t/T_0 \rfloor}$. In Fig.~\ref{Fig:dual_perform}, the stability of queue lengths for two selected nodes is depicted over the $T = 100$ time steps. It is evident that as the dual variables increase, the queue lengths at these nodes grow correspondingly. However, once the dual variables reach their optimal levels, the queue lengths stabilize at their most favorable values.

\begin{figure}[htp]
    \centering   
    \subfloat[\centering Abilene \label{10(a)}]{%
    \includegraphics[width=0.5\linewidth]{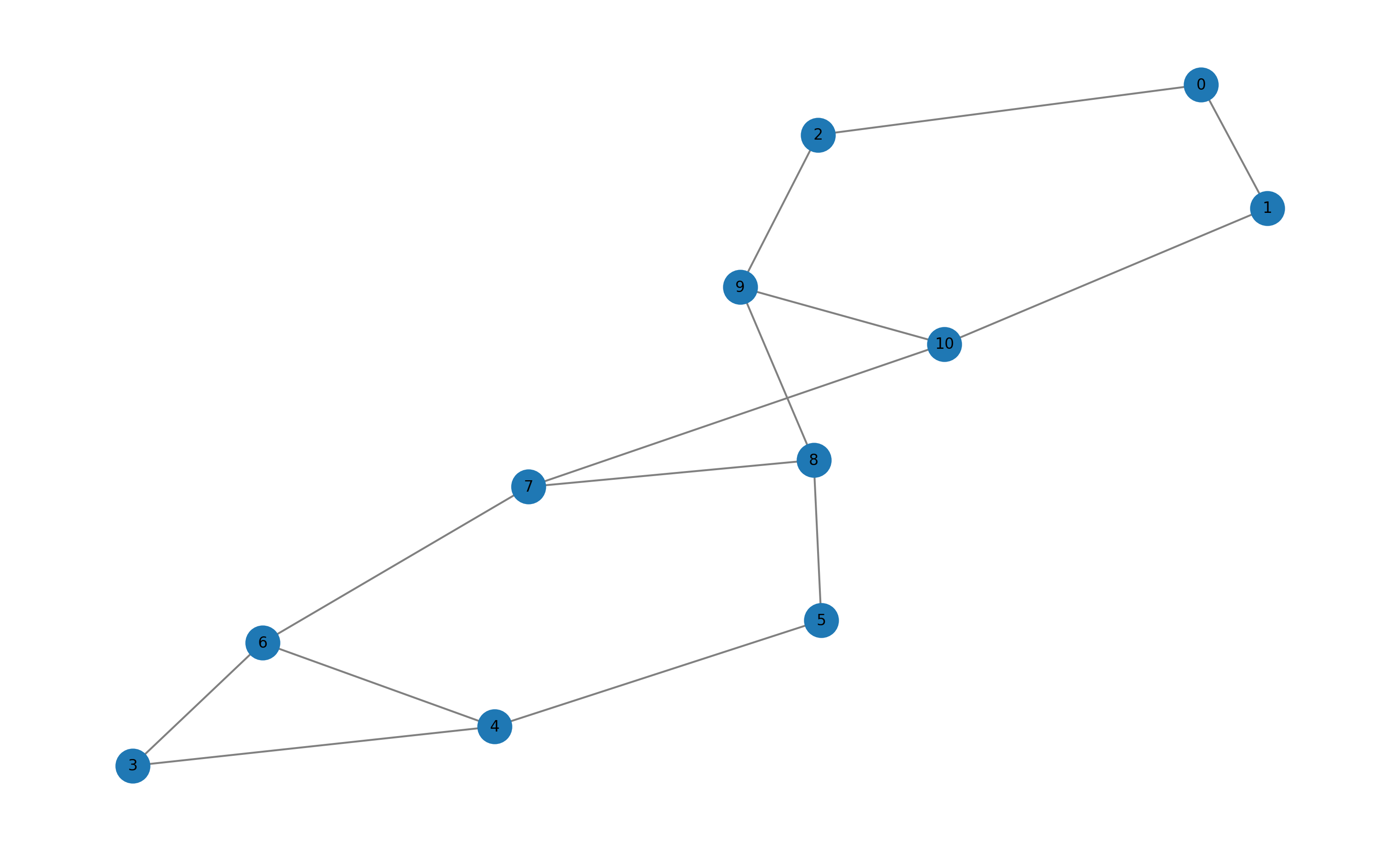}}
    \hfill
    \subfloat[\centering Tinet \label{10(b)}]{%
    \includegraphics[width=0.5\linewidth]{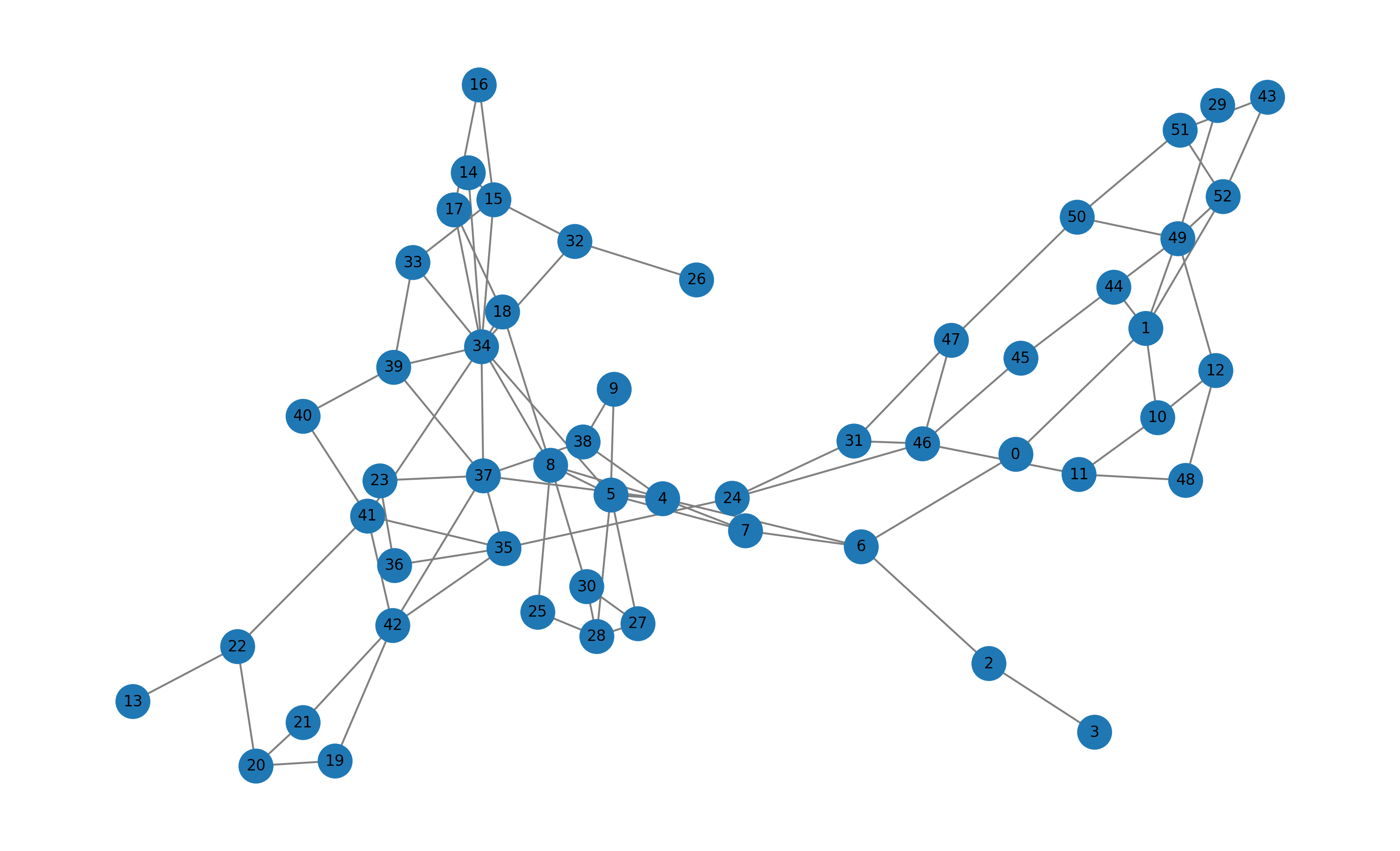}}
    \hfill
    \subfloat[\centering Sinet \label{10(c)}]{%
    \includegraphics[width=0.5\linewidth]{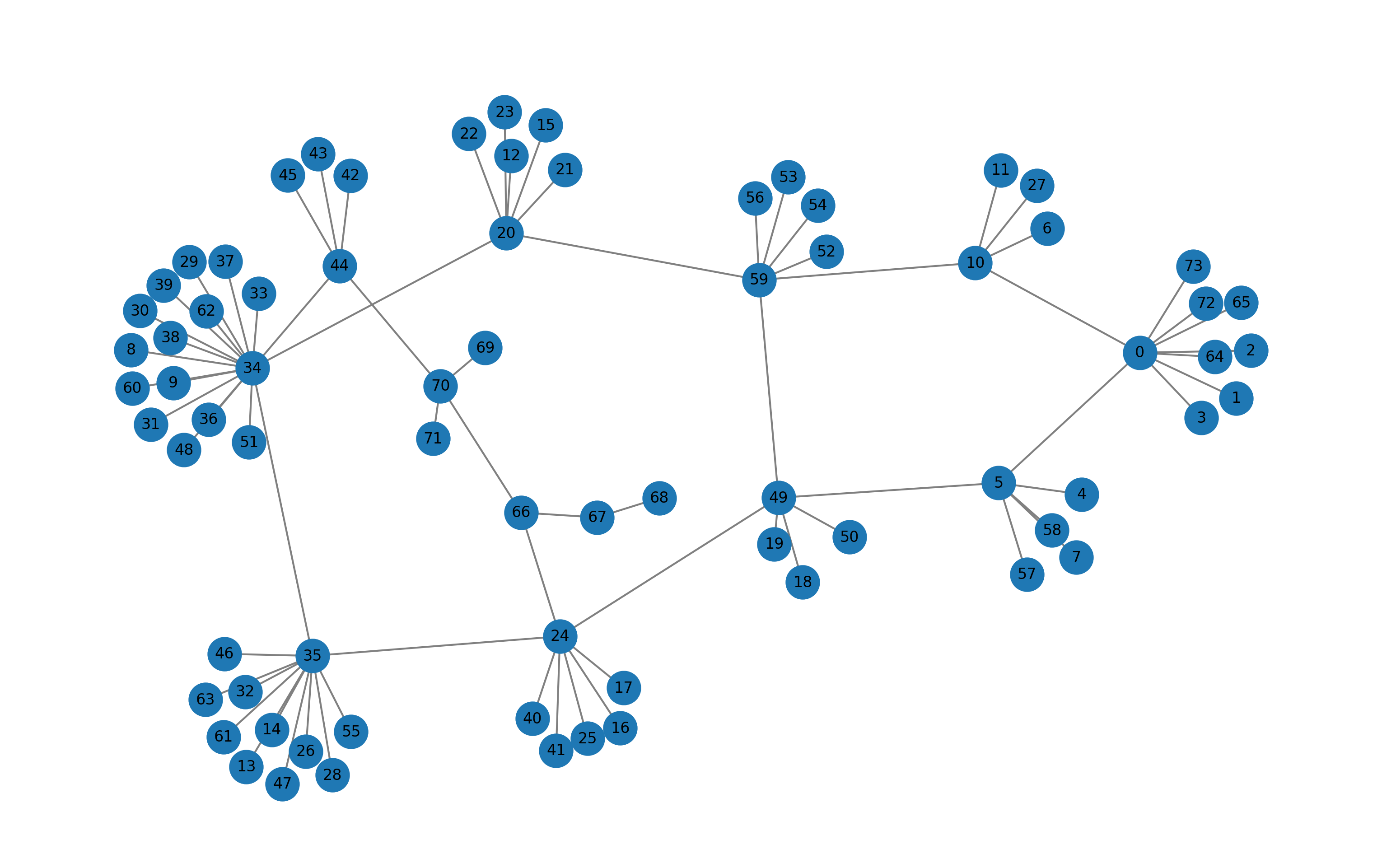}}
    \hfill
    \subfloat[\centering Interoute \label{10(d)}]{%
    \includegraphics[width=0.5\linewidth]{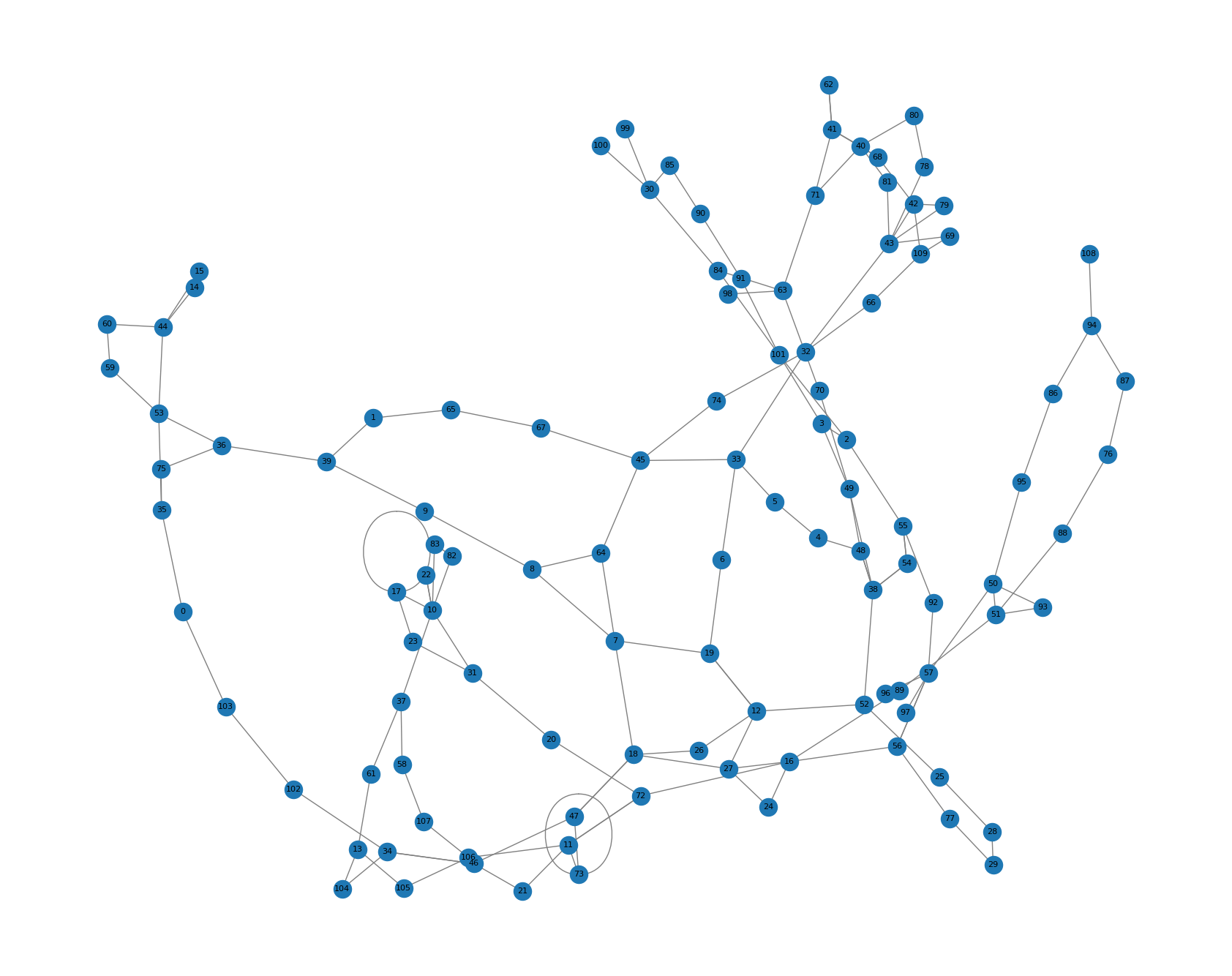}}
    \caption{Network topology graphs sourced from the Internet Topology Zoo dataset.}
    \label{Fig:real_topology}
\end{figure}

\subsection{Performance on Network Topology Datasets}
\begin{figure}[h]
    \centering   
    \includegraphics[width=3.5in]{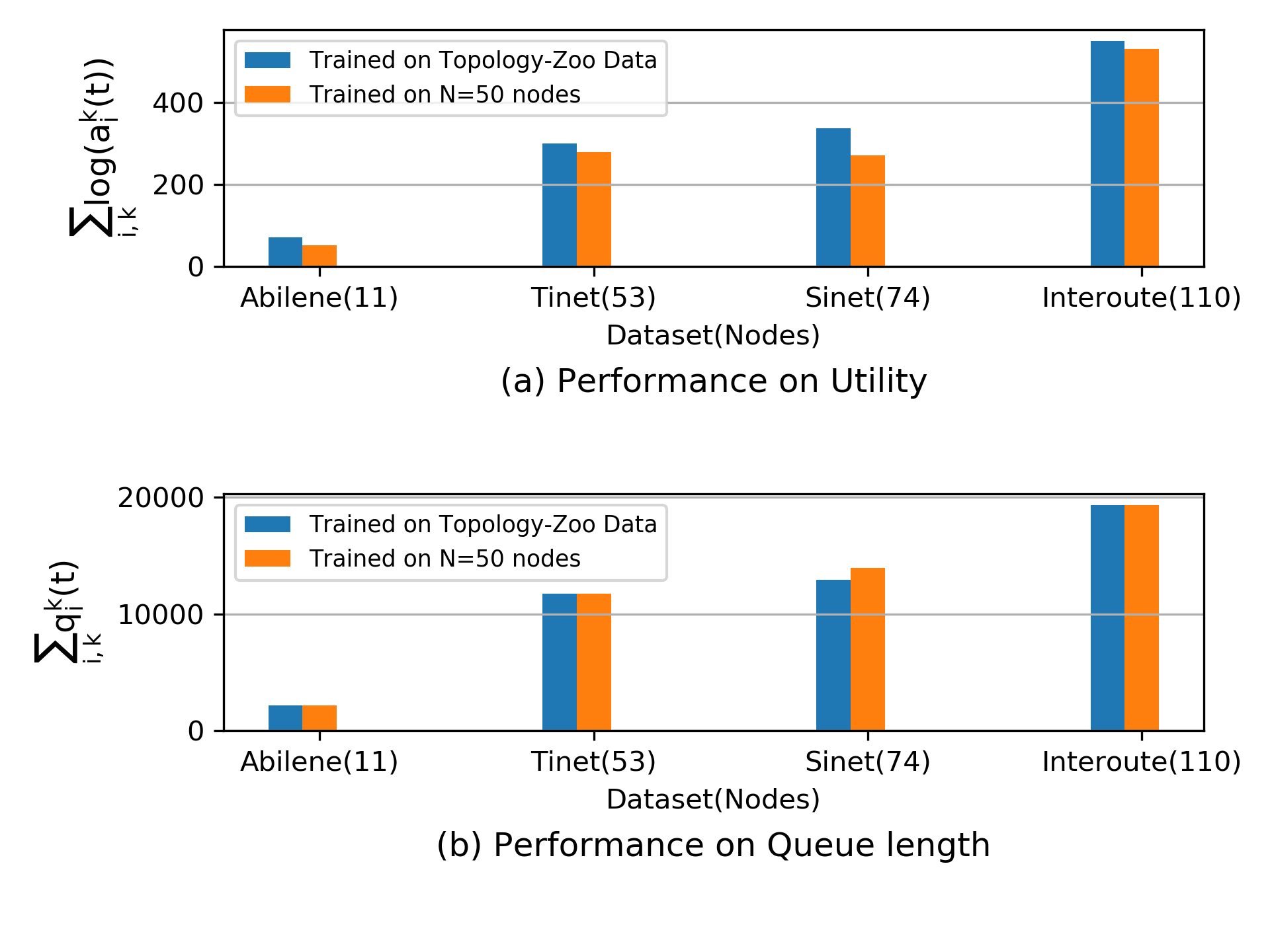}
    \caption{Evaluation of the state-augmentation-based routing algorithm on real-world network topologies.}
    \label{Fig:real_data}
\end{figure}

We further evaluated our proposed methodology on a range of real-world network configurations derived from the Internet Topology Zoo dataset \cite{knight2011internet}. Models were trained on four distinct network topologies, as illustrated in Fig.~\ref{Fig:real_topology}, with 4 flows. To test the versatility of the GNN architecture, a model trained on a 50-node network with 4 flows using randomly generated data was applied to these real-world networks. As shown in Fig.~\ref{Fig:real_data}, while models trained and tested on the same network topologies demonstrate strong performance, the transferred model from the 50-node random data network delivers comparable results. 
This observation demonstrates that our proposed model is generalized to effectively handle diverse datasets and network topologies.

\subsection{Performance on Real-Time Testbed}
Our final observation was to evaluate the validity of the trained GNN models on a real time wireless mesh network. We developed a wireless network testbed using Raspberry Pi's which served as the nodes of the network. Within the limits of experimental resources, we constructed wireless ad-hoc networks of size 8 nodes and 10 nodes for 3 flows in the laboratory of University of Pennsylvania. We evaluated the trained GNN models on these networks and transferred the model trained on 8 node network to the network with 10 nodes. As we can see in Fig.~\ref{Fig:real_data}, the model trained on a network with 8 nodes and 3 flows performs quite well as compared to the model which was already trained on the 10 node network. This showcases that the proposed state-augmentation algorithm can indeed be implemented in practice as well the transferability properties of GNN models to larger and unseen networks. 

\begin{figure}[h]
    \centering   
    \includegraphics[width=3.5in]{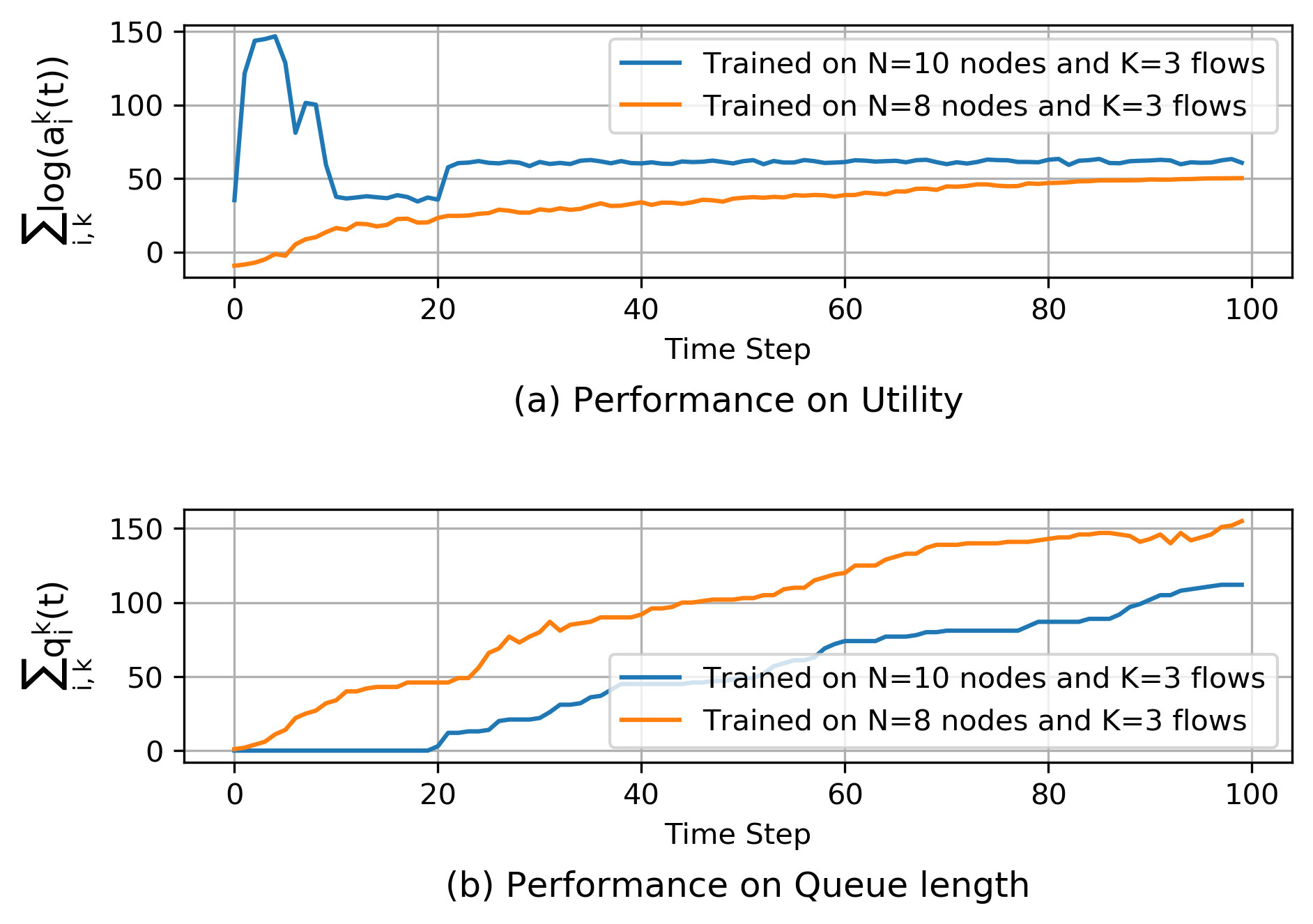}
    \caption{Transferability of model trained on 8 node network to a network with 10 nodes on a wireless communication testbed.}
    \label{Fig:rpi_data}
\end{figure}

\section{Conclusion} \label{sec:conclusion}
This study investigates the challenge of routing information packets in a wireless communication network, with the objective of optimizing a network-wide utility function while satisfying multiple constraints. 
Our experiments showed learning-based methods outperforming conventional optimization approaches, despite incurring initial training costs. However, unparameterized learning, while feasible and near-optimal, required infinite iterations and frequent re-optimization for each set of dual variables. 
In order to overcome these issues, we proposed a state-augmentation-based routing algorithm to address these challenges. Through experiments and analysis, we observed that our state-augmented learning framework effectively generates feasible and near-optimal routing decisions for a broad range of parameterizations. Additionally, we utilized Graph Neural Networks (GNNs) to implement the state-augmentation-based routing optimization, showcasing their exceptional stability and adaptability across various networks, which includes network topologies derived from real-world datasets and real world wireless network testbeds. 

\bibliographystyle{IEEEtran}
\bibliography{references}








\end{document}